\documentclass{article} 
\usepackage{iclr2022_conference,times}

\usepackage{amsmath,amsfonts,bm}

\def\eqref#1{equation~\ref{#1}}

\def\1{\bm{1}}

\DeclareMathAlphabet{\mathsfit}{\encodingdefault}{\sfdefault}{m}{sl}
\SetMathAlphabet{\mathsfit}{bold}{\encodingdefault}{\sfdefault}{bx}{n}

\usepackage{microtype}
\usepackage{graphicx}

\usepackage{subcaption}

\usepackage{booktabs} 
\usepackage{xcolor} 
\usepackage{pgfplots,pgfplotstable}

\usepackage{hyperref}
\usepackage{url}

\newcommand\boldhead[1]{\vspace{0.03in}\noindent\textbf{#1: }}

\usepackage{amsmath}
\usepackage{amssymb}
\usepackage{mathtools}
\usepackage{amsthm}
\usepackage{bm}
\usepackage[capitalize,noabbrev]{cleveref}

\definecolor{forestgreen}{rgb}{0.13, 0.55, 0.13}
\definecolor{fulvous}{rgb}{0.86, 0.52, 0.0}
\definecolor{glaucous}{rgb}{0.38, 0.51, 0.71}
\definecolor{lava}{rgb}{0.81, 0.06, 0.13}
\definecolor{buff}{rgb}{0.94, 0.86, 0.51}
\definecolor{chromeyellow}{rgb}{1.0, 0.65, 0.0}
\definecolor{brightube}{rgb}{0.82, 0.62, 0.91}

\hypersetup{
    pdfborderstyle={/S/U/W 1}, 
    linkbordercolor=red,       
    citebordercolor=green,     
    filebordercolor=magenta,   
    urlbordercolor=cyan        
}

\pgfplotsset{scaled y ticks=false, scaled x ticks=false}

\pgfplotstableread[col sep=comma]{data/sigma_sweep_final.csv}\sigmasweepdata
\pgfplotstableread[col sep=comma]{data/bgap_prior_loss_plot.csv}\bgappriorlossdata
\pgfplotstableread[col sep=comma]{data/agap_prior_loss_plot.csv}\agappriorlossdata

\theoremstyle{plain}

\theoremstyle{definition}

\theoremstyle{remark}

\usepackage[textsize=tiny]{todonotes}

\usepackage{siunitx}

\title{Generative Modeling for Low Dimensional Speech Attributes with Neural Spline Flows}

\author{Kevin J. Shih \thanks{equal contribution} \And Rafael Valle\footnotemark[1] \And Rohan Badlani \And João Felipe Santos \And Bryan Catanzaro\\
NVIDIA\\
\texttt{\{kshih\} at [affiliation] dot com} \\
}

\iclrfinalcopy 
\begin{document}

\maketitle

\begin{abstract}
Despite recent advances in generative modeling for text-to-speech synthesis, these models do not yet have the same fine-grained adjustability of pitch-conditioned deterministic models such as FastPitch and FastSpeech2. Pitch information is not only low-dimensional, but also discontinuous, making it particularly difficult to model in a generative setting. Our work explores several techniques for handling the aforementioned issues in the context of Normalizing Flow models. We also find this problem to be very well suited for Neural Spline flows, which is a highly expressive alternative to the more common affine-coupling mechanism in Normalizing Flows.
\end{abstract}

\section{Introduction}
Our work aims to bridge the gap between two families of text-to-speech (TTS) models: the deterministic but heavily factorized models, and the less factorized generative models. In the first group, works such as FastSpeech2~\citep{ren2020fastspeech}, FastPitch~\citep{lancucki2021fastpitch}, and Mellotron~\citep{valle2020mellotron}, conditioned not only on text, but also on pitch, and
sometimes energy. These models are capable of fine-tuning results for
applications such as cross-speaker pitch transfer, as well as
auto-tune like effects. Furthermore, while clean audio is hard to come by, acoustic features such as pitch are robust to noise and various distortions. This means that parts of these models responsible for modeling pitch distributions can incorporate large amounts of less-than-perfect training data with no loss in quality. However, being deterministic, there is a loss in realism as these models will never vary given the same prompt. On the other hand, we have generative models such as
GlowTTS~\citep{kim20glowtts}, GradTTS~\citep{popov2021grad}, and RADTTS~\citep{shih2021rad}. These models are capable of producing diverse samples
from the same prompt, but lack the aforementioned factorized benefits. What if we could have it all?



\begin{figure}[h]
\centerline{\includegraphics[height=0.1\textheight,width=1\columnwidth]{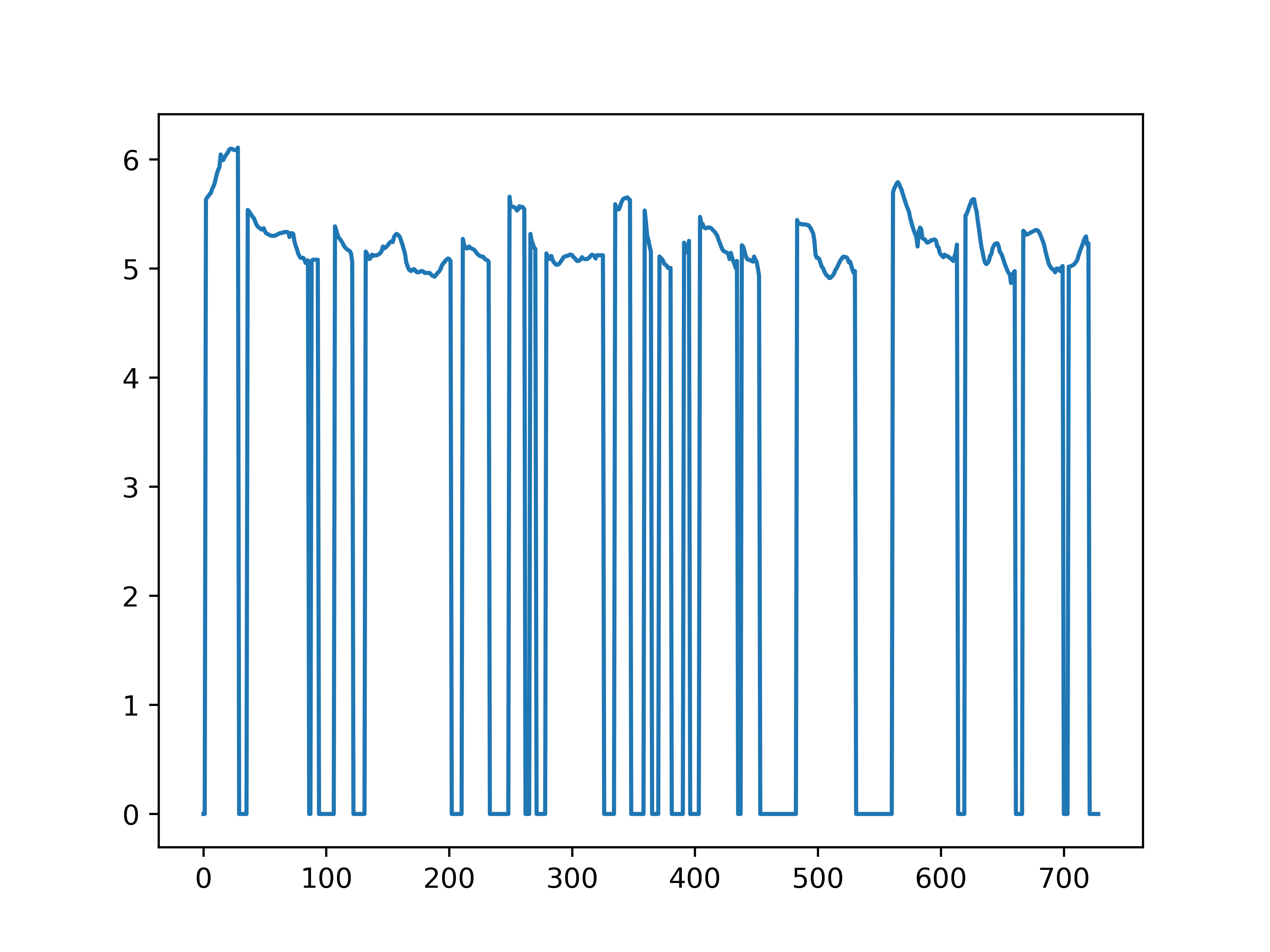}}
\caption[]{Example graph of fundamental frequency of a speech sample plotted against time in log space. Unvoiced regions are aperiodic and hence have no frequency data, depicted with a placeholder value of 0 in this graph.}
\label{fig:F0ex}
\end{figure}

In order to achieve the same level of factorization in a generative
context, one must be able to fit a generative model over fundamental frequency or pitch.
However, we can quickly see why this is a difficult task. In practice,
pitch information is represented by the fundamental frequency ($F_0$) of
a speaker's voice. Consider Figure~\ref{fig:F0ex}, which depicts the
$log(F_0)$ of a speech sample. We can identify the following issues upon inspection:

\boldhead{Low dimensionality} As with audio waveform data, fundamental
frequency is a 1D waveform and low dimensional data is
difficult to work with. For the same reason that audio waveform data
is first expanded to 80+ dimensional Mel Spectrograms using STFT, we
must either find a reasonable means to increase the dimensionality, or
find a model architecture that excels on low dimensional data.

\boldhead{Discontinuity}
Unlike audio waveform data, fundamental frequency data is discontinuous. It comprises segments of periodic voiced regions with valid fundamental frequency interleaved by aperiodic unvoiced segments with no fundamental frequency. How one handles the unvoiced regions is important because as far as the model is concerned, all inputs are valid inputs. As a direct result of the discontinuity issue, the variance in graph appears much more than it really is, due to the artificial transitions between the valid data and placeholder values.
A generative model attempts to map \emph{all} the variance in the graph to a Gaussian. If the model is not aware of the differences between these regions, one could end with unexpected spikes and dips in the middle of the spoken segments, leading to catastrophic audio artifacts.

Our work explores several techniques for handling the aforementioned issues, using normalizing flows as our framework for generative modeling. Our contributions are as follows:
\begin{itemize}
    \item We propose normalizing flow models that are aware of voiced/unvoiced segments, and demonstrate why this is critical.
    \item We compare various techniques for handling issues regarding $F_0$ data.
    \item We explore various model architectures, including both parallel and autoregressive normalizing flow models, as well as the use of neural spline flows in place of the standard affine-coupling.
\end{itemize}

\section{Related Works}
\boldhead{Feature Augmentation} Normalizing flows are known to be limited to homeomorphism-preserving transformations. Prior works such as~\citep{huang2020augmented} and ~\citep{chenVflow} propose the incorporation of additional dimensions by augmenting with random variables drawn from a different distribution. Similarly, ~\citep{dupont2019augmented}, proposes to append a vector of zeros to circumvent a similar limitation in Neural ODEs. \citep{kim2020softflow} tackles a similar problem of fitting normalizing flow models to map thin 3D structures to Gaussian priors. Instead of augmenting dimensions, they propose to dilate the data by directly adding noise, creating an augmented data distribution that is topologically closer to the target prior. Our work draws inspiration from these prior works, however our proposed approaches are more specialized to tackle the intricacies of discontinuous fundamental frequency information in speech.

\boldhead{Generative Text-to-Speech}
Several recent works have tackled the problem of generative text-to-speech modeling, wherein the goal of the resulting model is to sample diverse outputs given a single text prompt. Flowtron~\citep{valle2020flowtron} proposes an autoregressive-flow whereas GlowTTS~\citep{kim20glowtts}, FlowTTS~\citep{miao20flowtts} both propose a Glow-style~\citep{kingma2018} non-autoregressive flow model for sampling text-conditional mel-spectrograms. RADTTS \citep{shih2021rad} further extends the flow-based approaches by incorporating generative duration synthesis, replacing deterministic phoneme regression used in prior methods. More recently, likelihood models based on diffusion denoising probabilistic models~\citep{jeong2021diff,popov2021grad} have been considered as well. Our work operates in the same vein as these prior methods, but our target pitch and energy representations occur \emph{before} the mel-spectrogram stage, thereby allowing us to achieve diverse synthesis with the option of frame-level manual adjustment.

\boldhead{Generative F0 Modeling}
While most deep-learning based approaches have focused on deterministic regression approaches, there have been some generative autoregressive approaches, using RNNs to output stochastic states or Gaussian mixture densities ~\citep{wangDAR,wangMDN,wangVQDAR}. These approaches similarly separate the inference task into the voiced and unvoiced cases, which has been a commonly used formulation in parametric vocoder such as \citep{mccree1995mixed}. Our work builds upon these methods, but we additionally explore non-autoregressive (parallel) architectures in addition to being primarily focused on overcoming the known limitation of normalizing flows on low dimensional data.


\section{Methods}
To construct a generative model while maintaining fine-grained
adjustability over speech
characteristics such as pitch and energy, we focus on fitting
generative Normalizing Flow models over the
fundamental frequency ($F_0$) of audio segments, including voiced/unvoiced decisions, 
as well as the average energy per mel-spectrogram frame, referred to as energy for
brevity. This is in contrast to most existing generative TTS models,
which directly model the stochastic generation of mel-spectrogram
frames. While the methods we discuss apply to both $F_0$ and energy, we will
focus our discussion on the much more problematic $F_0$.

We will first give a brief overiew of Normalizing flows. Next, we will
discuss data preprocessing techniques for resolving $F_0$-related
issues in the context of Normalizing Flow frameworks, followed by a
discussion of model design.

\subsection{Normalizing Flows Overview}
As with most generative models, Normalizing Flows allows us to map our
data domain to a Gaussian distribution such that we can sample
new data points with ease. Let $X$ be the data domain (in our case,
$F_0$), and $Z$ be the normally-distributed latent space:
\begin{align}
  z \sim \mathcal{N}(0, I) \\
  x = G(z;\theta) 
\end{align}

Normalizing Flows are based on the change-of-variables formula, which
states that if $G(z;\theta)$ is invertible ($z = G^{-1}(x;\theta)$), we
can  then optimize for its parameters $\theta$ with exact MLE. We define $f_{Z}(z)$ and  $f_{X}(x)$ as the
probability density function in the latent (Gaussian) domain, and the
unknown pdf in the data domain respectively. The change of variables
gives us the following formula:
\begin{equation}
  \ln f_{X}(\mathbf{x}) = \ln f_{Z}(G^{-1}(\mathbf{x};\theta)) +
  \ln\bigg\lvert \det \frac{\partial G^{-1}}{\partial \mathbf{x}} \bigg\rvert
\end{equation}
Using substitution, it then follows that the following objective
function is the equivalent of exact maximum likelihood over the data:
\begin{equation}
  \hat{\theta}  = \arg \max_{\theta}  \ln f_{Z}(G^{-1}(\mathbf{x};\theta)) +
  \ln\bigg\lvert \det \frac{\partial G^{-1}}{\partial \mathbf{x}} \bigg\rvert
\end{equation}
Here, $\hat{\theta}$ are the parameters of a neural architecture
$G()$, carefully constrained to
guarantee invertibility.

As we are primarily interested in inferring $F_{0}$ and Energy in a
text-to-speech framework, $G()$ is further conditioned on
$\Phi_{text}$, which is a matrix specifying temporally-aligned text
information. As such, we have the invertible model $G(\cdot;\theta,
\Phi_{text})$, which models the conditional distribution of $F_{0}$ or
Energy given temporally aligned text.

\subsection{Increasing Data Dimensionality}
\label{sec:datadim}
We first discuss several potential methods for tackling the low
dimensionality problem with $F_0$ and energy. Low dimensionality is
particularly problematic in Normalizing Flows due to the bijectivity
constraint. The bijectivity constraint limits us to homeomorphic
mappings between $X$ and $Z$, which can be extremely limiting in low dimensions. Furthermore, we cannot simply throw in
fully-connected layers to project our data to a higher dimension, as
these transformations would not be invertible. Instead, we rely on the use of data grouping, auxiliary
dimensions, and approximately invertible transformations.

\boldhead{Grouping}
Grouping is a technique commonly used in fitting normalizing flow models to time-domain-based data. Let $X: x^{1}, x^{2}, x^{3} \ldots x^{t} \ldots x^{T}$ be our domain data comprising a sequence of $T$ real values. The superscript indicates the sequence index. Each $x \in X$ is a $D$-dimensional data point. We can group every $N$ consecutive data points together in a non-overlapping fashion, thereby resulting in $T/N$ $ND$-dimensional data points. For example, let $N=2$, then $X^{\prime}=(x^{1}, x^{2}), (x^{3}, x^{4}) \ldots (x^{t}, x^{t+1}) \ldots (x^{T-1}, x^{T})$. While it is possible to widen indefinitely, we found that increasingly larger group sizes result in reduced variability in sampling. As such, one should go with the minimum group size that one can get away with. A group size of 2 was sufficient for $F_0$, but a group size of 4 was necessary for modeling the energy distribution.

\boldhead{Auxiliary Dimensions}
Another way to increase the dimensionality without breaking
bijectivity is to tack on auxiliary dimensions, which we simply
discard later during inference. We find that the an approximation of
the local derivative at every time step $x^{t}$ to be simple yet effective.

Continuing from the grouping example, we arrive at:
$X^{\prime}=(x^{1}, \frac{\partial x^{1}}{\partial{t}}, x^{2}, \frac{\partial
x^{2}}{\partial{t}}), \ldots (x^{T-1}, \frac{\partial x^{T-1}}{\partial{t}},
x^{T},\frac{\partial x^{T}}{\partial{t}})$. The derivative values work
similarly to grouping, but also provide additional context information
summarizing the relationship between the current group and the
adjacent ones. We compute the approximate local derivative by
computing the centered-difference at each timestep $t$ and scaling as
necessary for model stability:
\begin{equation}
  \frac{\partial x^{t}}{\partial{t}} =
  \frac{(x^{t}-x^{t-})+(x^{t+1}-x^{t})} {\kappa}
\end{equation}

An alternative to the above is to use a set of basis functions to project the signal to a higher dimension. Following FastSpeech2~\citep{ren2020fastspeech}, we also consider the continuous wavelet
transformation. Similar to STFT, continuous wavelet transform (CWT)~\citep{cwtsuni} can be used to convert a $F_0$ contour into a time-frequency representation,
giving us more dimensions to work with. We compare this against the centered-difference auxiliary features as described above. Please check section~\ref{sec:cwt} for further implementation details and discussion.

\subsection{Filling In The Holes}
\label{sec:datafill}
In addition to giving the models more dimensions to work with, it is
still necessary to fill in the gaps between voiced regions of the $F_0$
data. Using a constant-value filler is problematic as it is very
difficult for the Normalizing Flow model to map a segment of
zero-variance data to a Gaussian. The most straightforward solution is linear interpolation, as
part of the CWT transformation. However, we are also interested in
potential solutions that avoid hallucinating values within the same
range of valid data. This way, we can still identify unvoiced and
voiced segments of the data from the graph.

\begin{figure}
 \centering
     \begin{subfigure}[b]{\columnwidth}
         \centering
         \includegraphics[height=0.1\textheight,width=\textwidth]{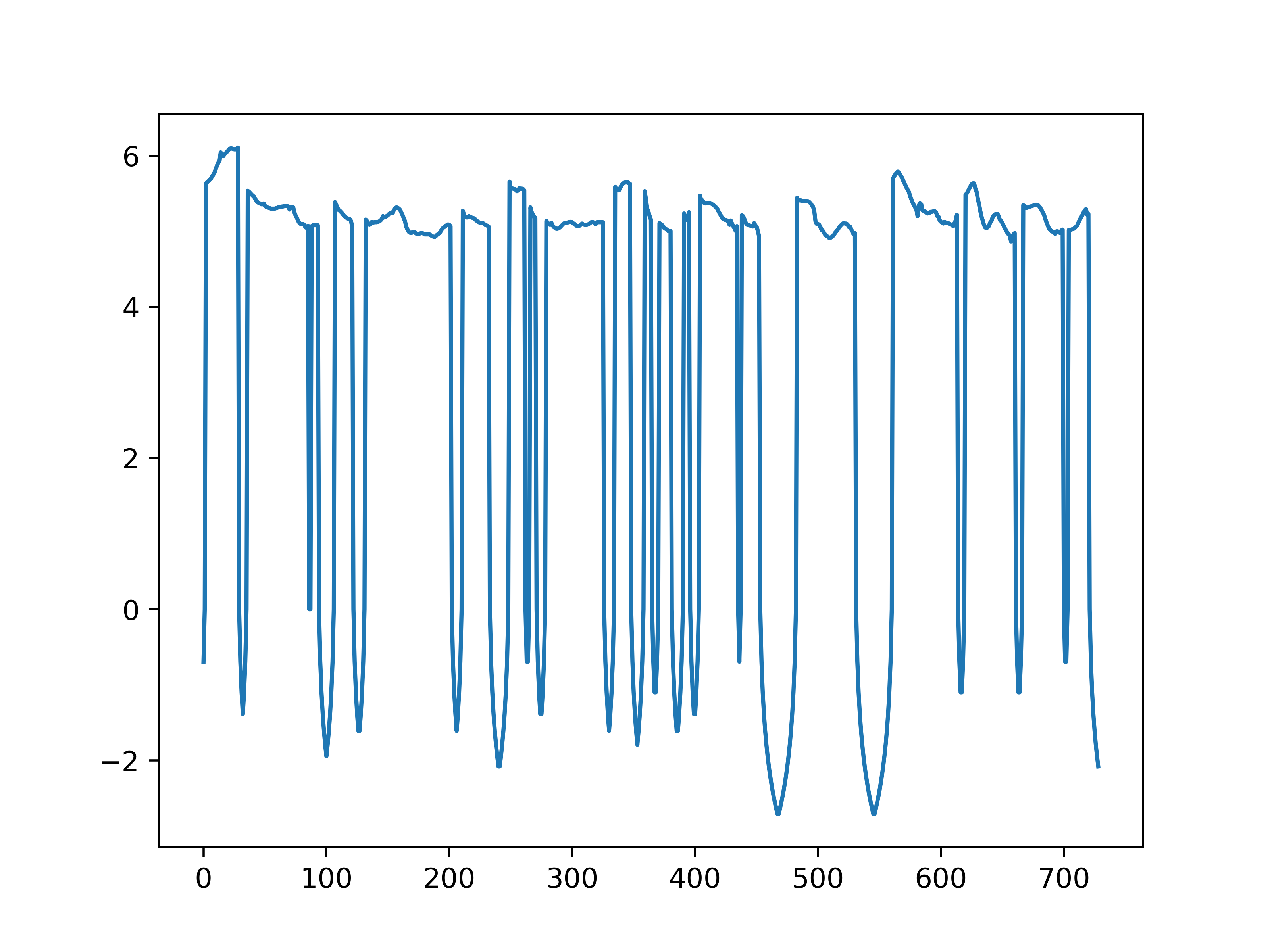}
          \caption{log distance transform filler}
         \label{fig:f0dtx}
     \end{subfigure}
     \hfill
     \begin{subfigure}[b]{\columnwidth}
         \centering
         \includegraphics[height=0.1\textheight,width=\textwidth]{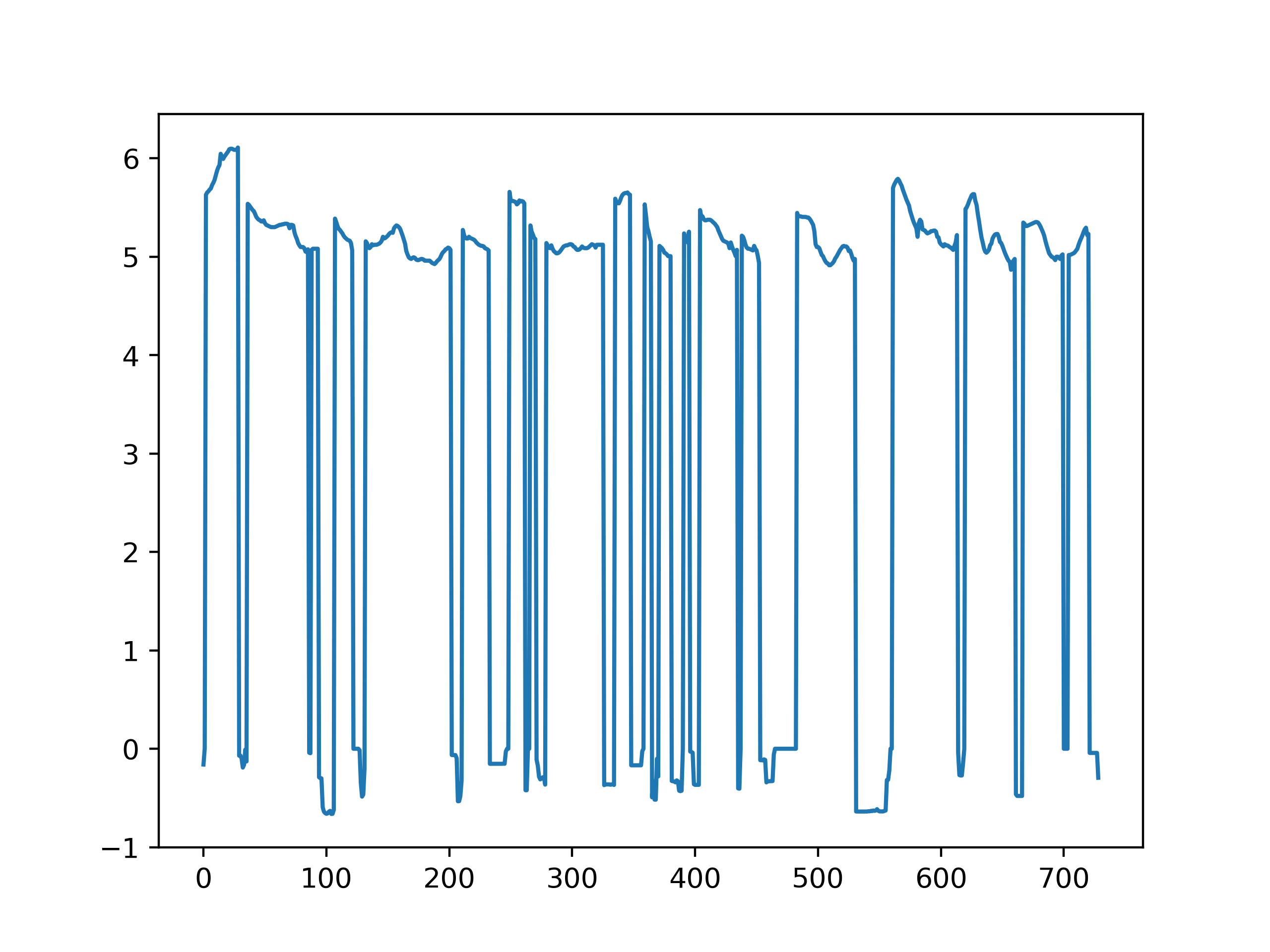}
             \caption{learned bias filler}
         \label{fig:f0unvb}
     \end{subfigure}
    \caption{$F_{0}$ data with filler data between voiced segments to avoid long stretches of constant values. We consider both input text agnostic (\ref{fig:f0dtx}) and input text dependent (\ref{fig:f0unvb}) variants.}
    \label{fig:f0filler}
\end{figure}

\boldhead{Log-Distance Transform Filler}\label{sec:holefilling}
We consider the use of a distance transform to fill in the values for
the unvoiced region, where the value at every time step is the minimum
distance to the next voiced segment. We use the log of the distance
transform, else the resulting value of
the centered-difference auxiliary dimensions would be a constant
value of $\pm 1$. Finally, we negate the values to avoid overlapping
with the voiced $F_0$ values. An example can be seen in Fig.~\ref{fig:f0dtx}.

\boldhead{Unvoiced Bias}
Recall that our $F_{0}$ modeling task is part of a TTS pipeline, and
thus the curvature is conditioned on the corresponding text (phoneme) at each time step.
One potential drawback of the distance transform is that it is agnostic
to the underlying phoneme sequence. As such, we also consider a 
different solution -- one that learns to infer negative offsets in the 
unvoiced regions conditioned on the phoneme sequence. 
An example can be seen in Fig.~\ref{fig:f0unvb}.
We will elaborate  on the implementation details in section \ref{sec:agap}.

\subsection{Model Architectures}
The focus of this work is on modeling the distribution of $F_{0}$
and energy in speech, conditioned on temporally aligned text. As we
are working within the framework of a TTS pipeline, we
assume we have the following components available:
\begin{figure}
  \centering
  \includegraphics[width=0.5\columnwidth]{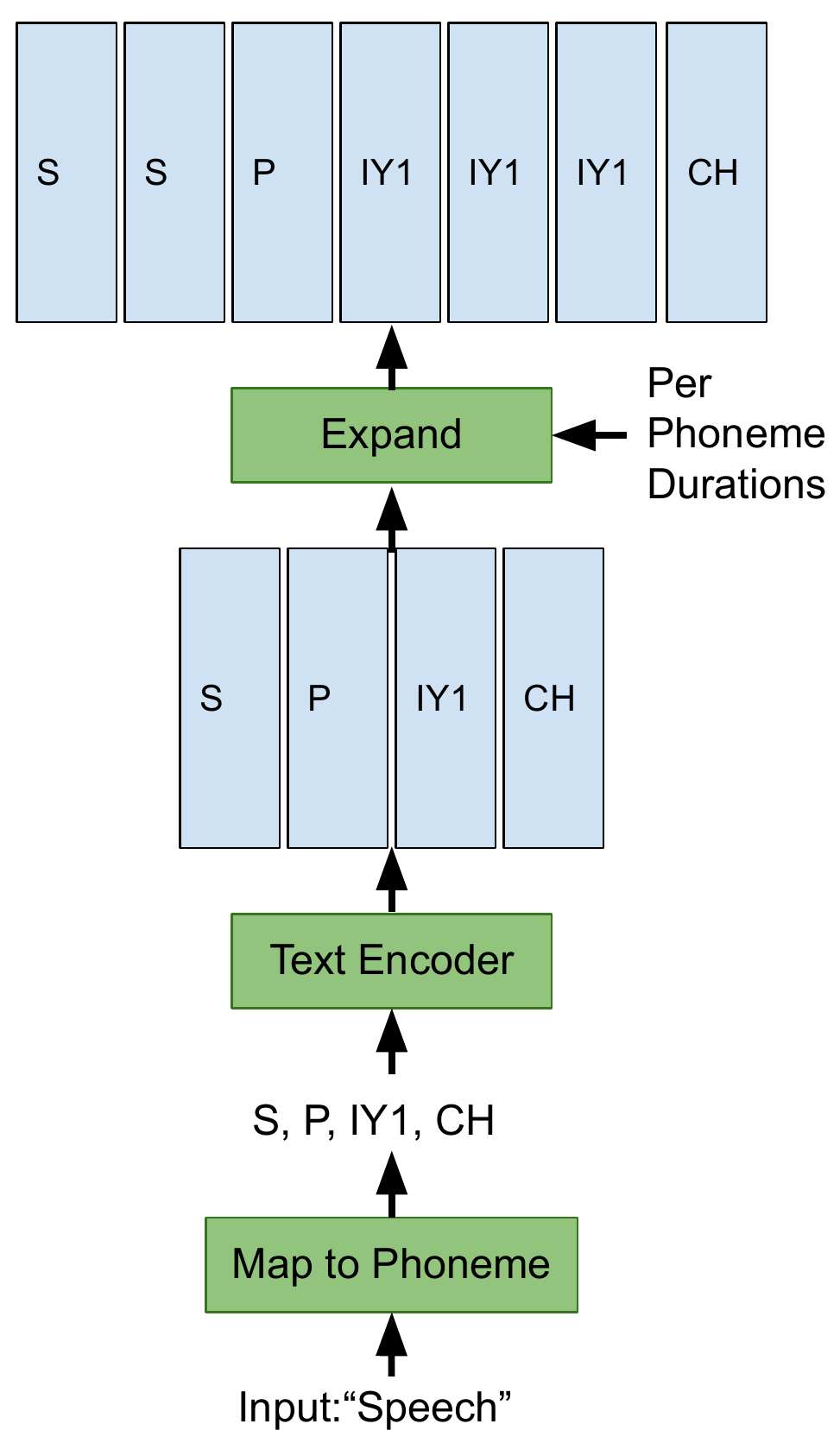}
  \vspace{-1em}
  \caption{Construction of $\Phi_{text}$ from input ``speech''. Text
    is mapped to phonemes, which are then encoded into individual
    feature vectors. Each vector is then replicated based on the
    specified duration of the respective phoneme. The topmost matrix
    is the resulting $\Phi_{text}$}
    \label{fig:phitext}
\end{figure}

\boldhead{Timed Text Representation} $\Phi_{text}$ is a $C \times T$ matrix that contains both
textual information and the timing of individual phonemes within said
text. Each slice $\Phi^{t}_{text}$ gives us a $C$-dimensional feature
vector representing the textual information at time $t$. We visualize this in Fig~\ref{fig:phitext}. The goal of the model is to fit the distribution $P(X|\Phi_{text})$, where $X$ corresponds to either $F_{0}$ or energy ($\mathcal{E}$).

\boldhead{Pitch and Energy Conditioned Mel-Decoder} Typically in TTS
models, a decoder first maps textual inputs to mel-spectrograms, which
are then converted to waveform audio using a vocoder. As with works
such as FastPitch~\citep{lancucki2021fastpitch} and FastSpeech2, 
we assume our mel decoder is further conditioned on pitch ($F_{0}$) 
and energy information:
\begin{equation}
D(\Phi_{text}, F_{0}, \mathcal{E}) \to X_{Mel}
\end{equation}

To achieve this, we use a reimplementation of the RADTTS~\citep{shih2021rad} architecture, modified to be further conditioned on $F_0$ and $\mathcal{E}$. Interestingly, we found that conditioning on $F_0$ and $\mathcal{E}$ explain away most of the perceived variations in the resulting samples coming out of the decoder, despite RADTTS being a generative model in its own light. As such, the onus of producing diverse synthesis is now assigned to the generative $F_0$ and $\mathcal{E}$ models.

Importantly, the RADTTS alignment mechanism (in a similar fashion to GlowTTS), provides the necessary audio-text timing information necessary to construct $\Phi_{text}$. 

\boldhead{Vocoder} Finally, we have the vocoder, which takes in
$X_{mel}$ and gives us our final waveform output for audio. We use
Hifi-GAN~\citep{kong2020hifi} for this purpose. The vocoder choice has
little relevance to the focus of this work.

Given the above components, we consider two architectures for modeling
pitch and energy given $\Phi_{text}$: Glow-style~\citep{kingma2018}
bipartite model and one based on inverse-autoregressive
flows~\citep{kingma2016improved}.

\subsubsection{Bipartite Flow Model}
The proposed bipartite model is structurally similar to the
bipartite flow component used in RADTTS~\citep{shih2021rad}, with
similarities to the one in GlowTTS~\citep{kim2020glow}. However, the
focus on modeling low dimensional discontinuous data poses additional
challenges that warrant further architectural changes.

\begin{figure}
    \centering
    \includegraphics[width=0.7\columnwidth]{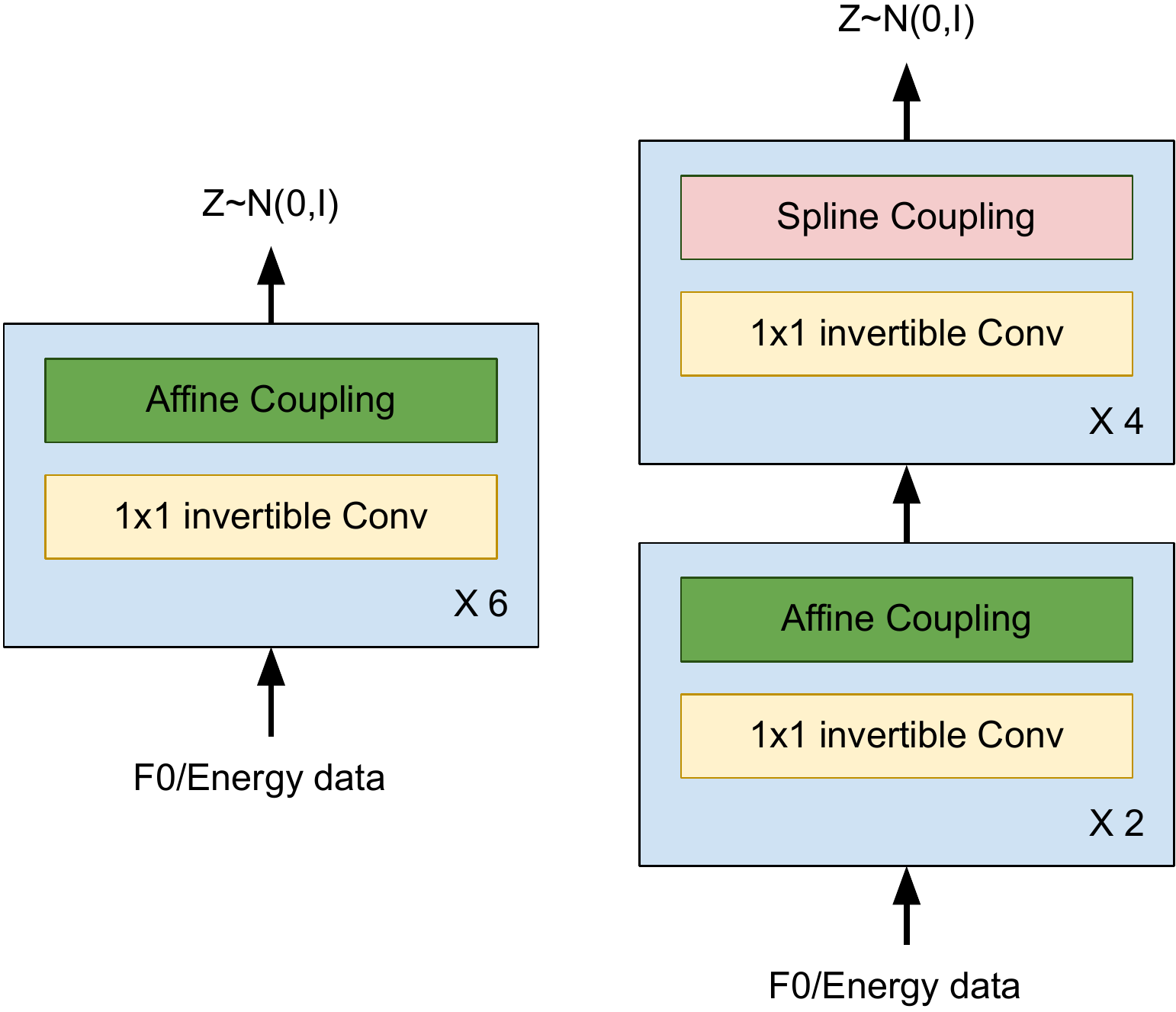}
    \vspace{-1em}
    \caption{On the left depicts a standard Glow-style generative model using flow steps comprising pairs of 1x1 invertible convolutions and affine coupling blocks. We consider up to 6 steps total for our work. The left side depicts our hybrid spline-affine architecture, replacing the affine coupling blocks in the 4 steps nearest $Z$ with more powerful spline transformations, but retaining 2 affine blocks for added stability.}
    \label{fig:bgaparch}
\end{figure}

\boldhead{Data Format} Attempting to fit directly to raw $F_{0}$ and
$\mathcal{E}$ data is incredibly unstable, thereby necessitating the
data transformations in Sections \ref{sec:datadim} and
\ref{sec:datafill}. The training is unstable in that not only does the
likelihood loss have difficulty converging on training data, but that
the \emph{posterior distribution of the training data struggles to conform
to a standard normal}. The latter indicates that the
model cannot find a reasonable mapping between $f_{X}$ and $f_{Z}$,
rendering the model unusable. The use of grouping and auxiliary
dimension are strictly necessary for model stability, whereas
additional improvements in output quality are attained through proper handling of unvoiced segments.

We train separate bipartite models for $F_{0}$ and $\mathcal{E}$ respectively.
We use a grouping of size 2 for $F_{0}$, combined with either
centered-difference auxiliary features or CWT. The resulting
dimensionality at each time step is 4 and 24 respectively. We use a
grouping of size 4 with centered-difference auxiliary features for
$\mathcal{E}$, giving us a dimensionality of $4 \times 2 = 8$. We
found the model to be unstable with group size below 4. Difficulties
in modeling $\mathcal{E}$ possibly stem from energy being only weakly
correlated with the underlying text in $\Phi_{text}$, and more
strongly correlated with nearby values.

\boldhead{Voiced-Aware Context}\label{sub:voicedawarecontex}
The quality of our $F_0$ prediction plots improves with each of the
aforementioned strategies. However, catastrophic errors still happen . The
patterns for $F_{0}$ in the voiced and unvoiced segments are bimodal;
they cannot be mixed. Failures occur when the model performs a
misplaced transition between modes, such as in the middle of a voiced
segment. Errors such as these result in outlier spikes and dips in the
generated $F_{0}$ curvature, resulting in catastrophic audio artifacts. Examples are shown in section \ref{sec:bipartitequal}.
 
 Fortunately, one can easily construct a voiced/unvoiced classifier on which to condition the flow model's output. 

\begin{figure}
    \centering
    \includegraphics[width=\columnwidth]{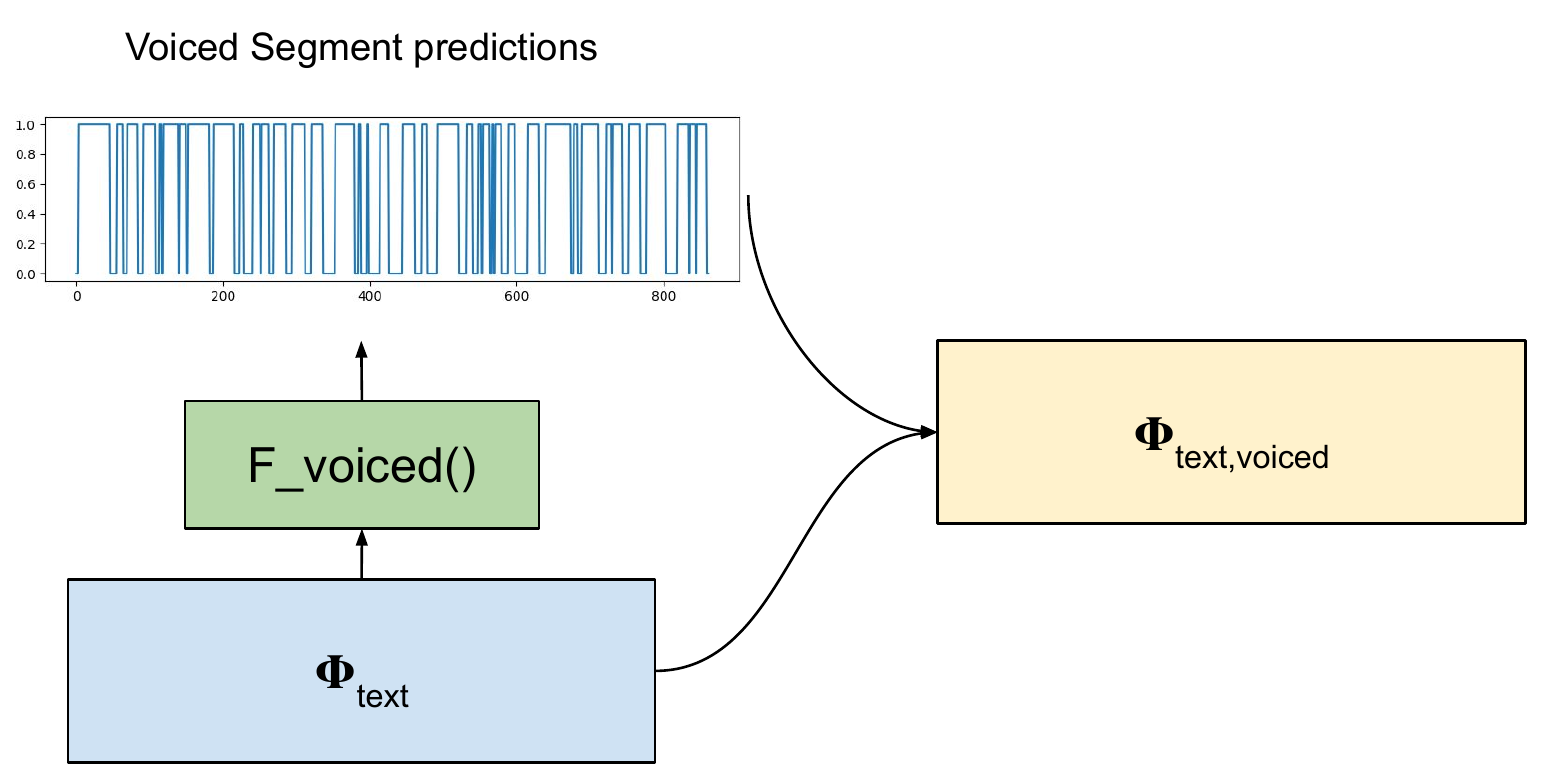}
    \caption{Abstract depiction of incorporating voiced-segment awareness into $\Phi_{text}$. A classifier infers a binary voiced-segment mask from $\Phi_{text}$, the result of which is merged back into $\Phi_{text}$ using the formula in (\ref{eq:voicedmerge}). Ground-truth voice masks are used to supervise during training.}
    \label{fig:voicedmerge}
\end{figure}
Conditioning on explicit
voiced/unvoiced region labels prevents the flow model from performing
misplaced transitions. We construct such a classifier $V=F(\Phi_{text})_{voiced}$ where $F_{voiced}$ is a regression function that predicts at
every time step $t\in T$ whether we are dealing with a voiced or an
unvoiced time step. Here, we assume $V$ is a binary vector of length
$T$ such that $V \in \{0,1\} ^{1\times T}$. We achieve this by
thresholding regressed values at 0.5.

Given our predicted voiced/unvoiced mask $V$ and our context matrix
$\Phi_{text}$, one could simply concatenate them and condition the
flow model on the concatenated result. However,
concatenation runs the risk of the model choosing to ignore
single-channel $V$, which
has a negligible vector norm compared to each slice in
$\Phi_{text}$, set to 512 channels in our implementation. Instead, we learn separate affine transformations for
voiced and unvoiced regions to apply directly on $\Phi_{text}$ to get
$\hat{\Phi}_{text}$, similar to conditional instance normalization
used in style transfer literature~\citep{Dumoulin2017}. Let $V^{t}$ be
the binary voiced/unvoiced indicator at timestep $t$. $\bm{s}_{voiced}$,
$\bm{b}_{voiced}$, $\bm{s}_{unvoiced}$, and
$\bm{b}_{unvoiced}$ are $C$-dimensional embedding vectors, matching the
dimensions in $\Phi_{text}$. They represent the learned scale ($\bm{s}$)
and bias ($\bm{b}$) for voiced and unvoiced regions. We perform the
voiced-conditional affine transformation of $\Phi_{text}$ as follows:

\begin{align}
  \alpha^{t} &= \sigma(V^{t}\bm{s}_{voiced} + (1-V^{t})\bm{s}_{unvoiced})\\
  \beta^{t} &= \tanh(V^{t}\bm{b}_{voiced} +
                (1-V^{t})\bm{b}_{unvoiced})\\
  \hat{\Phi}_{text,voice} &= \alpha^{t}\Phi^{t}_{text} + 0.01\beta^{t}
  \label{eq:voicedmerge}
\end{align}

The result is a per-channel affine transformation over $\Phi_{text}$, with different
affine parameters for voiced and unvoiced regions, allowing us to condition on timed text \emph{and} whether a time segment is voiced. We can use the
ground truth voiced/unvoiced area mask during training, and the output
of $F_{voiced}$ during inference.

\subsubsection{Autoregressive Flow Model}\label{sec:agap}
\begin{figure}
    \centering
    \includegraphics[width=\columnwidth]{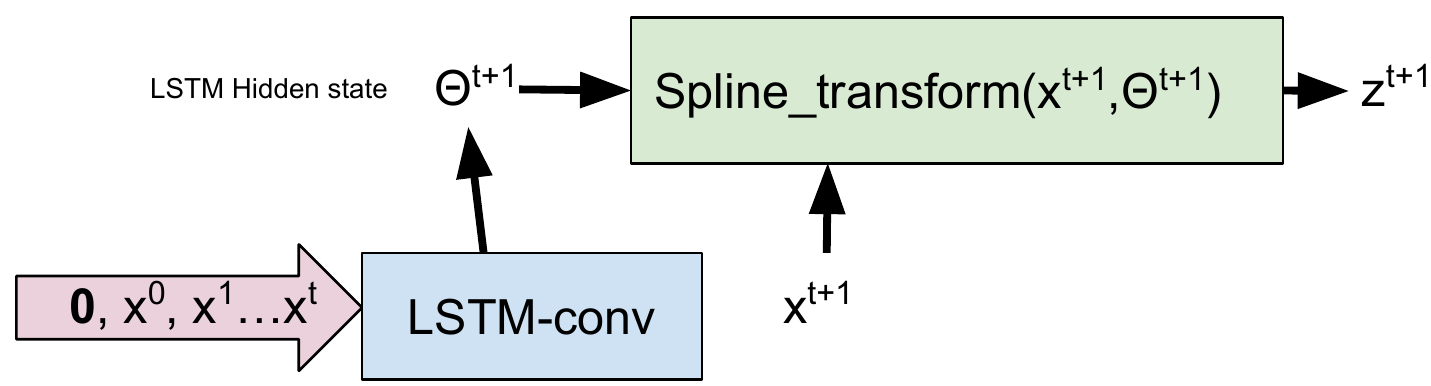}
    \caption{Abstract depiction of the autoregressive flow architecture. A recurrent architecture processes data in a specified direction, generating the transformation parameters for $x^{t+1}$ conditioned on $x^{0}$ to $x^{t}$. A constant value 0 is prepended to the data to ensure the process is invertible. Our work explores the use of splines in place of affine transforms in the autoregressive transformation process.}
    \label{fig:agaparch}
\end{figure}

Our autoregressive flow model for modeling $F_{0}$ and $\mathcal{E}$
is based on the bidirectional autoregressive flow architecture used in
Flowtron~\citep{valle2020flowtron,papamakariosNeurips17}. As with before, we train separate
models for $F_{0}$ and $\mathcal{E}$. A high level description of how the model works can be seen in Figure~\ref{fig:agaparch} with a more detailed description of the forward and inverse process given in section \ref{sec:agapfi}.

Following the implementation in~\citep{valle2020flowtron}, the neural network architecture comprises a 2-layer LSTM followed by a final non-linear projection over the hidden state to produce the affine or spline transformation parameters. The full model comprises two such models: one running from left-to-right, and a second identical model running from right-to-left on the output of the first, similar to a bi-directional LSTM.

The autoregressive transformation is more expressive than the bipartite transformation as the corresponding Jacobian matrix of the autoregressive transform is much more dense compared to that of a bipartite model~\citep{ping2020waveflow}. This difference is especially significant on low-dimensional data such as what we are attempting in this work.

\boldhead{Learning the Unvoiced Bias} 
Similarly to the bipartite model, the autoregressive model is unstable on raw $F_{0}$ data due to unvoiced phonemes. Drawing inspiration from mixed excitation synthesizers~\citep{mccree1995mixed} in which unvoiced sounds were represented as a distribution over frequency band and width, we learn a negative bias term for each unvoiced phoneme. 

Let $\phi_{text}$ represent the sequence of duration agnostic phoneme embeddings in an utterance. We design a regression function $b_{unvoiced} = -ReLU(F(\phi_{text})_{bias})$ that predicts a per-phoneme negative or zero bias given phoneme embedding. The voiced/unvoiced mask described in \ref{sub:voicedawarecontex} operates at a frame basis, hence we first align the unvoiced bias with the $F_0$ contour by using ``ground truth" phoneme durations. Then we apply the voiced mask to the unvoiced bias and finally bias the $F_0$ contour. We optimize $F_bias$ by backpropping through the $F_0$ of the conditional decoder. As this is ReLU-based, we still end up with some zeros in the unvoiced regions. Nevertheless, the autoregressive model worked will with this setup.

Further improvements in handling the voiced and unvoiced region transitions come from incorporating the same voiced-aware context as described in the bipartite model description. Without it, the model relies heavily on it its hidden states, which are far less reliable.

Worth noting that while the learned bias worked well for the autoregressive model, the phoneme-agnostic distance transform filler resulted in unnatural-sounding $F_{0}$ contours. Conversely, the bipartite model was somewhat more stable using the distance transform than the learned bias, as the latter may still have zeros in the unvoiced region. We speculate that the difference in behavior is due to the distinct differences in learned data associations between an autoregressive process and a conv-net-style bipartite model.


\subsubsection{Improved Model Fitting with Neural Splines}
Both normalizing flow architectures we examine rely on coupling-based
transforms. The most commonly used implementation is the
affine-coupling function, which splits the data along the channel
dimension and uses the one split to infer a set of affine parameters
for the other:
\vspace{-0.2em}
\begin{align}
  x \to_{split} x_{a},x_{b}\\
  f(x_{b})\to D,\beta\\\label{eq:parampred}
  x^{*}_{a} = Dx_{a}+\beta\\
  concat(x^{*}_{a}, x_{b})\to x^{*}
\end{align}

Here, $D$ is a diagonal matrix of inferred scaling values to simplify
the inversion and log-determinant calculation process, which are $1/D$
and the sum of the diagonal terms on $\log D$ respectively. As $x_{b}$ is unaltered, we can recover $D$ and $\beta$ to invert
the transformation on $x_{a}$.

One potential limiting factor of affine coupling is that the set of
transformation parameters is uniform for all possible values in
$x_{a}$. This is relevant to our case because, as previously stated,
our $F_{0}$ data representation is bimodal. It would not make sense to
use the same affine parameters for both voiced and unvoiced values.

For this reason, we consider Neural Splines as an alternative to
affine coupling. Neural Splines, as originally proposed by
~\citep{muller2019neural} and further extended by
~\citep{durkan2019neural}, replace the affine function with a
monotonic piecewise polynomial function. This function is monotonic
and its input and output are bounded, thereby allowing for easy
invertibility. Importantly, the piecewise spline transformation
comprises of a series of connected polynomial functions. A
value in $x_{a}$ will receive different treatment based on which
spline's bin it falls into. Critically, this allows the model to learn
multi-modal transformations necessary for multi-modal inputs.

We use piecewise quadratic splines for our work, simply adjusting the
parameter predictor in (\ref{eq:parampred}) to output the spline
parameters instead of affine. One limitation of neural spline coupling
layers is that the input and output are bounded. While it is customary
to simply to use the identity function for anything outside of bounds,
this can be problematic if too much (or sometimes all) of the data ends
up outside of bounds.  For the bipartite architecture, we replace the
affine coupling transform in the 4 out of 6 flow steps closest to the
latent space with neural splines, leaving the 2 steps closest to the
training data as affine coupling. Having the neural splines deal
directly with the bimodal $F_{0}$ data can potentially lead to some
instabilities. The
two simpler, but also more stable affine coupling flow steps serve to
massage the data slightly for the spline functions. The autoregressive
architecture comprises only two affine coupling blocks, one for each
direction in the bidirectional procedure. We replace each with neural
splines, carefully scaling the input data to fall within the spline's bounds.

Our experiments will later demonstrate that the introduction of neural
splines greatly improve the model's ability to map the training data
to the standard normal prior.

\section{Experiments}
\label{sec:experiments}
We conduct our experiments on the LJSpeech dataset\citep{ljspeech17}. The following
section includes ablation studies on various design choices, as well
as comparisons against deterministic baselines. All experiments, 
we use ``ground-truth" phoneme durations obtained from the RADTTS alignment framework to avoid introducing additional variables into the study. Our experiments focus on the quality of resampled $F_0$ and energy, with timing held as a constant. FastPitch (FP) uses a checkpoint available provided by the author~\citep{lancucki2021fastpitch}. Our FastSpeech2 (FS2)~\citep{ren2020fastspeech} is based on an third-party open-source implementation 
from \citep{fs2github}, modified and replicated to the best of our ability. 

When full-audio synthesis is necessary, synthesized features from our models are decoded using a reimplementation of the RADTTS mel spectrogram decoder, modified to further condition on $F_0$ and $\mathcal{E}$. A full study on its acoustic feature fidelity is provided in section~\ref{appendix:acousticfidelity}.

To go from mel spectrograms to waveform, all models share a single HiFi-GAN~\citep{kong2020hifi} checkpoint trained on LJSpeech as provided by the authors\citep{hifigangithub}. We do not fine-tune HiFi-GAN (HG) on synthesized mel-spectrograms because it would turn HG into a mel-spectrogram enhancer and introduce confounding factors into the analysis. For brevity, the bipartite model will be referred to as \textbf{BGAP} (bipartite generative attribute predictor) and the autoregressive architecture \textbf{AGAP}. We use the acronym RADTTS to refer to \citep{shih2021rad}, RADDAP to refer to RADTTS with deterministic attribute predictors, BGAP and AGAP to refer to bipartite and autoregressive flow models, whereas ablated model names are specified in the legend of Fig.\ref{fig:mse_global}.


\subsection{$F_0$ and $\mathcal{E}$ Distribution Evaluation}
We attempt to quantitatively evaluate how well our models are able to accurately reproduce the true distribution of the low-level acoustic features we are modeling. 

\subsubsection{Splines Improve Fit to Prior}
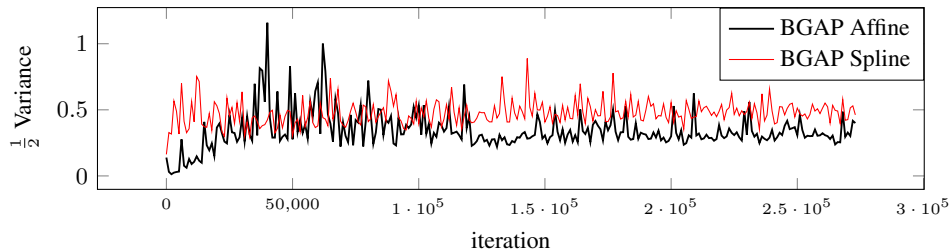
\begin{figure}
    \centering
    \begin{tikzpicture}[scale=1.0]
  \begin{axis}[
  height=4cm,
  width=0.9\linewidth,
       xlabel= {iteration},
       ylabel= {$\frac{1}{2}$ Variance},
      legend cell align={left},
      ylabel near ticks,
      xlabel near ticks,
      xlabel style={font=\small},
      ylabel style={font=\small},
      yticklabel style = {font=\small},
      xticklabel style = {font=\tiny},
      y tick label style={/pgf/number format/fixed,/pgf/number format/precision=3},
    legend style={font=\small},
    legend columns=1,
    legend entries={BGAP Affine,BGAP Spline},
    legend style={at={(1.0,1.0)},anchor=north east},,
    ]
    \addplot [line width=0.25mm,color=black] table [x=iter, y=affineploss, col sep=comma] {\bgappriorlossdata};
    \addplot [color=red] table [x=iter, y=splineploss, col sep=comma] {\bgappriorlossdata};
  \end{axis}
\end{tikzpicture}\vspace{-1em}
    \caption{Comparison of per-batch variance on training data between an affine and spline BGAP model for $F_0$. Plot values represent $\frac{1}{2}$ of the variance assuming a zero-mean distribution, thus 0.5 indicates a standard normal. We observe that latent distribution for the affine model struggles to conform to a standard normal even at convergence.}
    \label{fig:priorloss}
\end{figure}

Normalizing flows fit a transformation, mapping the input data distribution to 
a (usually) a standard normal. We observed that the distribution of the projected latent values per training batch would quickly conform to a standard normal, whereas unstable models would not. Figure~\ref{fig:priorloss} tracks the value $\frac{1}{2}||z||_2$ taken from the normalizing flow loss function. This value also corresponds to $\frac{1}{2}$ the variance of a zero-mean Gaussian, which means we should expect it to stabilize at exactly $0.5$. We observe that in the bipartite case, the introduction of neural splines significantly improve the rate at which the projected training data conform to a standard normal. The baseline model using only affine transforms never conform to a standard normal even at convergence. However, the same comparison on the autoregressive architecture resulted in little to no difference, as both affine-only and spline architectures had no trouble transforming the distribution.


%
\subsubsection{Statistical Moments}
Following \citep{ren2020fastspeech}, we compare statistical moments $\mu_n$ of the ground truth $F_0$ distribution, represented as midi notes~\citep{moog1986midi}  ($m = 12 \times \log_2(\frac{f}{440}) + 69$) against the synthetic pitch distribution from several models. Whereas the first two statistical moments provide a measure of what an average note is and how much variation there is, the third statistical moment highlights if notes lower than the mean are more frequent than notes higher than the mean, while the fourth moment highlights how often outlier notes are present and how high they are.

For these experiments, we use $\sigma_{F_0}=1$ to sample from all generative $F_0$ predictors and $\sigma_{\mathcal(E)}=1$ and $\sigma_{\mathcal(E)}=0.3$ for the autoregressive and the bipartite generative models. All models use voiced-aware contxet (Vpred). BGAP models use centered-difference first-order auxiliary features (FoF) in all cases except when continuous wavelet transforms (CWT) are used. We also ablate the use for distance transform filler data (DTx) for BGAP models. Our results in Table~\ref{tab:f0moms} show that with our method we effectively model the statistical moments without sacrificing quality. Our autoregressive flow model achieves the highest similarity with the target distribution in 4 out of 4 statistical moments, notably outperforming our internal deterministc RADDAP, which uses the same decoder and vocoder.

We hypothesize that the high kurtosis values on BGAP spline models are an artifact from instabilities due to the high $\sigma$ value, specially given the relatively small second moment. On the other hand, a comparison of the affine and spline BGAP models (rows 3 and 4) makes it evident that by adding the distance transform we are able to improve the third and fourth moments.

\begin{table}[!ht]
\small
\begin{center}
\caption{Statistical moments computed over midi notes. Closer to the GT value is better. Model numbers correspond to the same numbering as used in the Fig.\ref{fig:mse_global} legend.}
    \label{tab:f0moms}
    \begin{tabular}{c c|c|c|c|c}
        & Model   &  $\mathcal{\mu}_{1}$ & $\mathcal{\mu}_{2}$ & $\mathcal{\mu}_{3}$ & $\mathcal{\mu}_{4}$\\
        \hline
        & GT	 	& 55.47	            & 4.06  	    & 0.36	 	      & 0.50\\
        & HG	 	& 55.49	            & 4.00  	    & 0.26	 	      & -0.16\\
        \hline
        1& BGAP-AVDF	& 55.79	            & 3.53	        & -1.10	          & 18.30\\
        2& BGAP-SVC		& 56.41	            & 3.32	        & -0.04	          & -0.05\\
        3& BGAP-SVDF	& 55.63             & 3.07          & -6.67           & 196.01\\
        4& BGAP-SVF	 	& 55.63	            & 3.21	        & -10.81	      & 396.56\\
        5& \textbf{AGAP-SV}		& \textbf{55.56}	& \textbf{4.08}	& \textbf{0.33}	  & \textbf{0.70}\\
        6& AGAP-AV      & 55.27             & 4.89          & 0.52            & 1.06\\
        7& AGAP-SVC     & 56.00             & 3.92          & -0.09           & 0.05\\
        8& RADDAP 		& 56.36	            & 2.90	        & 0.04	 	      & 0.28\\
        9& FS2	 	    & 53.06	            & 6.86	        & -0.81           & -0.47\\
        10& FP		    & 55.15	            & 4.37	        & -1.30	          & 3.05\\
        \hline
    \end{tabular}
\end{center}
 \vspace*{-0.5\baselineskip}
\end{table}
\subsubsection{Sample Error}
We now look at mean squared error with respect to the ground truth on the LJS 100 validation set. We evaluate Voiced $F_0$ Error (VFE): the $F_0$ mean-squared error limited to the voiced segments of the data. Next we have the predicted energy error (ENR): the MSE between predicted energy and ground truth. Finally, we have Voicing Decision Errors (VDE)\citep{Nakatani2008}: mean-absolute error between predicted and ground truth voiced-region masks. As generative models do not predict the best sample everytime, we synthesize 30 samples per utterance to cover the distribution of errors given the same utterance. We use $\sigma_{F_0}=1$ for both models, and  $\sigma_{\mathcal(E)}=1$ and $\sigma_{\mathcal(E)}=0.3$ for AGAP and BGAP respectively. Our primary deterministic baseline is RADDAP. While we include results for FP and FS2, their mel spectrogram decoders have lower acoustic feature fidelity, see Appendix ~\ref{appendix:acousticfidelity}, and thus cannot be directly compared against.

\begin{figure}[h]
    \begin{center}
    \includegraphics[width=1.\linewidth]{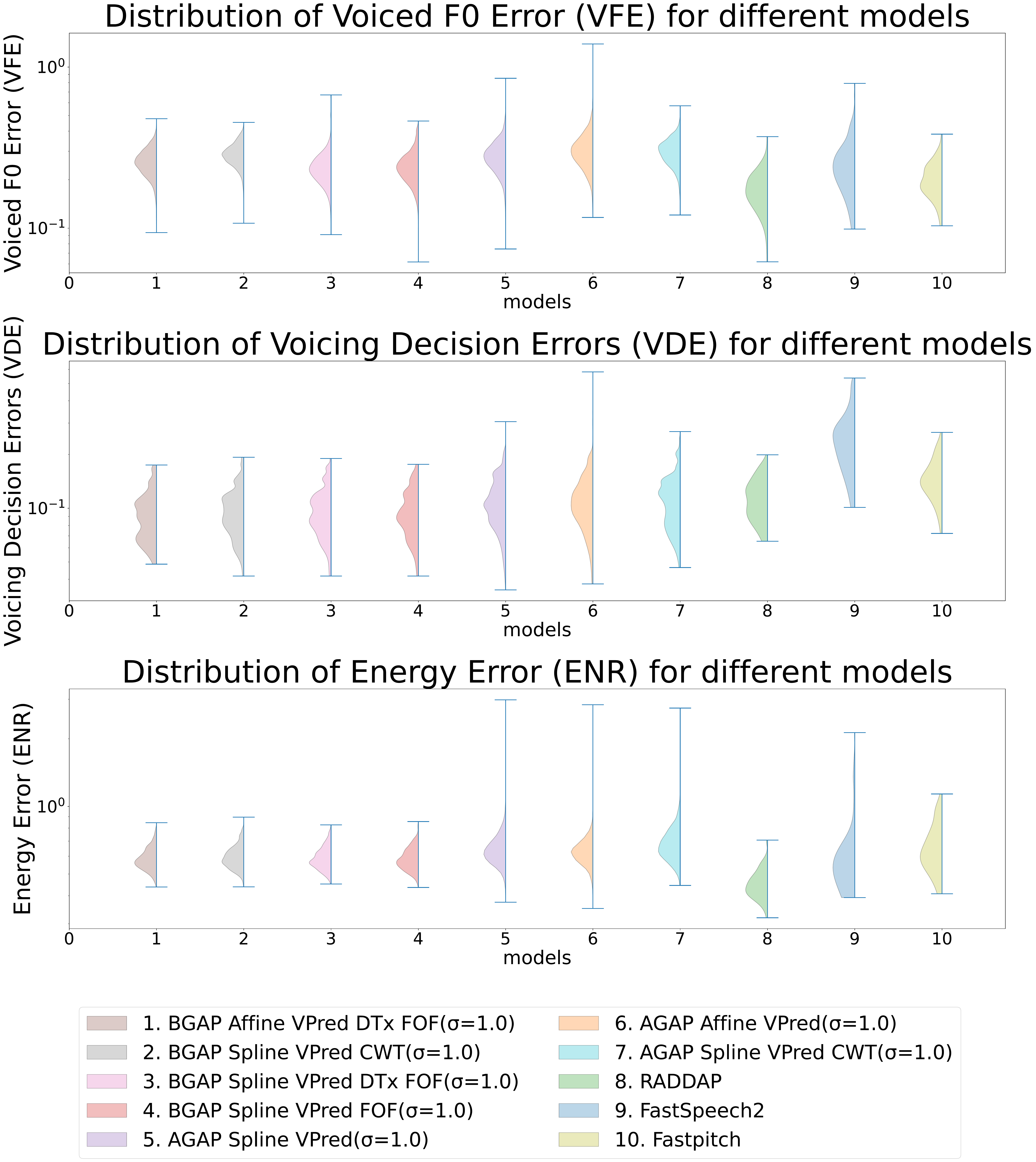}
    \vspace{-2em}
    \caption{Distribution of VFE, VDE and ENR between ground truth and synthesized samples. A comparison of the  distribution of FDE in deterministic and generative models highlights that generative modelling of $F_0$ is possible without loss in quality. In addition, the VDE error distributions suggest our context-aware context decreases voicing decision errors. The vertical axis is in $log$ space.}
    \label{fig:mse_global}
    \end{center}
\vspace{-2em}
\end{figure}

Figure \ref{fig:mse_global} provides  violin plots of the distribution of VFE, VDE and ENR computed over a set of models, with the vertical axis in $log$ space. 
By comparing models AGAP models 5 and 6, we can see the effect of replacing affine couplings layers with spline coupling layers. Incorporating splines considerably lowers the error distribution compared to the affine-coupling baseline.

The plot on the second row shows that our models in general have lower average VDE errors compared to other models, most notably our internal RADDAP baseline. We hypothesize this is due to the voiced-aware context we proposed. We also hypothesize that the relatively large error outliers in ENR for AGAP models compared to come from the difference in $\sigma_{\mathcal(E)}$ between AGAP and BGAP models.

\newpage
\section{Conclusion}
\label{sec:conclusion}
To the best of our knowledge, our work is the first to propose an explicit generative model for $F_0$ and Energy. We resolve the issue of generative modelling of low dimensional speech attributes by proposing solutions using both autoregressive and bipartite normalizing flow models.
Stability is achieved through voiced/unvoiced segment awareness, as well as several techniques for handling issues regarding discontinuities and zero derivative regions in $F_0$ data. Furthermore, we identify spline coupling layers as a powerful invertible transformation, particularly well suited for our task. 

\newpage
\bibliography{main}

\begin{thebibliography}{34}
\providecommand{\natexlab}[1]{#1}
\providecommand{\url}[1]{\texttt{#1}}
\expandafter\ifx\csname urlstyle\endcsname\relax
  \providecommand{\doi}[1]{doi: #1}\else
  \providecommand{\doi}{doi: \begingroup \urlstyle{rm}\Url}\fi

\bibitem[Chen et~al.(2020)Chen, Lu, Chenli, Zhu, and Tian]{chenVflow}
Jianfei Chen, Cheng Lu, Biqi Chenli, Jun Zhu, and Tian Tian.
\newblock {VF}low: More expressive generative flows with variational data
  augmentation.
\newblock In Hal~Daumé III and Aarti Singh (eds.), \emph{Proceedings of the
  37th International Conference on Machine Learning}, volume 119 of
  \emph{Proceedings of Machine Learning Research}, pp.\  1660--1669. PMLR,
  13--18 Jul 2020.
\newblock URL \url{https://proceedings.mlr.press/v119/chen20p.html}.

\bibitem[Chien(2022)]{fs2github}
Chung-Ming Chien.
\newblock Fastspeech2.
\newblock \url{https://github.com/ming024/FastSpeech2}, 2022.

\bibitem[Dumoulin et~al.(2017)Dumoulin, Shlens, and Kudlur]{Dumoulin2017}
Vincent Dumoulin, Jonathon Shlens, and Manjunath Kudlur.
\newblock A learned representation for artistic style.
\newblock \emph{ICLR}, 2017.
\newblock URL \url{https://arxiv.org/abs/1610.07629}.

\bibitem[Dupont et~al.(2019)Dupont, Doucet, and Teh]{dupont2019augmented}
Emilien Dupont, Arnaud Doucet, and Yee~Whye Teh.
\newblock Augmented neural odes.
\newblock \emph{Advances in Neural Information Processing Systems},
  32:\penalty0 3140--3150, 2019.

\bibitem[Durkan et~al.(2019)Durkan, Bekasov, Murray, and
  Papamakarios]{durkan2019neural}
Conor Durkan, Artur Bekasov, Iain Murray, and George Papamakarios.
\newblock Neural spline flows.
\newblock \emph{Advances in Neural Information Processing Systems},
  32:\penalty0 7511--7522, 2019.

\bibitem[Huang et~al.(2020)Huang, Dinh, and Courville]{huang2020augmented}
Chin-Wei Huang, Laurent Dinh, and Aaron Courville.
\newblock Augmented normalizing flows: Bridging the gap between generative
  flows and latent variable models.
\newblock \emph{arXiv preprint arXiv:2002.07101}, 2020.

\bibitem[Ito \& Johnson(2017)Ito and Johnson]{ljspeech17}
Keith Ito and Linda Johnson.
\newblock The lj speech dataset.
\newblock \url{https://keithito.com/LJ-Speech-Dataset/}, 2017.

\bibitem[Jeong et~al.(2021)Jeong, Kim, Cheon, Choi, and Kim]{jeong2021diff}
Myeonghun Jeong, Hyeongju Kim, Sung~Jun Cheon, Byoung~Jin Choi, and Nam~Soo
  Kim.
\newblock Diff-tts: A denoising diffusion model for text-to-speech.
\newblock \emph{arXiv preprint arXiv:2104.01409}, 2021.

\bibitem[Kawahara et~al.(2005)Kawahara, Cheveign{\'e}, Banno, Takahashi, and
  Irino]{kawahara2005nearly}
Hideki Kawahara, Alain~de Cheveign{\'e}, Hideki Banno, Toru Takahashi, and
  Toshio Irino.
\newblock Nearly defect-free f0 trajectory extraction for expressive speech
  modifications based on straight.
\newblock In \emph{Ninth European Conference on Speech Communication and
  Technology}, 2005.

\bibitem[Kim et~al.(2020{\natexlab{a}})Kim, Lee, Kang, Lee, and
  Kim]{kim2020softflow}
Hyeongju Kim, Hyeonseung Lee, Woo~Hyun Kang, Joun~Yeop Lee, and Nam~Soo Kim.
\newblock Softflow: Probabilistic framework for normalizing flow on manifolds.
\newblock \emph{Advances in Neural Information Processing Systems}, 33,
  2020{\natexlab{a}}.

\bibitem[Kim et~al.(2020{\natexlab{b}})Kim, Kim, Kong, and Yoon]{kim2020glow}
Jaehyeon Kim, Sungwon Kim, Jungil Kong, and Sungroh Yoon.
\newblock Glow-tts: A generative flow for text-to-speech via monotonic
  alignment search.
\newblock \emph{Advances in Neural Information Processing Systems}, 33,
  2020{\natexlab{b}}.

\bibitem[Kim et~al.(2020{\natexlab{c}})Kim, Kim, Kong, and Yoon]{kim20glowtts}
Jaehyeon Kim, Sungwon Kim, Jungil Kong, and Sungroh Yoon.
\newblock Glow-tts: A generative flow for text-to-speech via monotonic
  alignment search.
\newblock In \emph{Advances in Neural Information Processing Systems},
  2020{\natexlab{c}}.

\bibitem[Kingma \& Dhariwal(2018)Kingma and Dhariwal]{kingma2018}
Durk~P Kingma and Prafulla Dhariwal.
\newblock Glow: Generative flow with invertible 1x1 convolutions.
\newblock In S.~Bengio, H.~Wallach, H.~Larochelle, K.~Grauman, N.~Cesa-Bianchi,
  and R.~Garnett (eds.), \emph{Advances in Neural Information Processing
  Systems}, volume~31. Curran Associates, Inc., 2018.

\bibitem[Kingma et~al.(2016)Kingma, Salimans, Jozefowicz, Chen, Sutskever, and
  Welling]{kingma2016improved}
Durk~P Kingma, Tim Salimans, Rafal Jozefowicz, Xi~Chen, Ilya Sutskever, and Max
  Welling.
\newblock Improved variational inference with inverse autoregressive flow.
\newblock \emph{Advances in neural information processing systems},
  29:\penalty0 4743--4751, 2016.

\bibitem[Kong(2022)]{hifigangithub}
Jungil Kong.
\newblock Hifi-gan: Generative adversarial networks for efficient and high
  fidelity speech synthesis.
\newblock \url{https://github.com/jik876/hifi-gan}, 2022.

\bibitem[Kong et~al.(2020)Kong, Kim, and Bae]{kong2020hifi}
Jungil Kong, Jaehyeon Kim, and Jaekyoung Bae.
\newblock Hifi-gan: Generative adversarial networks for efficient and high
  fidelity speech synthesis.
\newblock \emph{Advances in Neural Information Processing Systems}, 33, 2020.

\bibitem[{\L}a{\'n}cucki(2021)]{lancucki2021fastpitch}
Adrian {\L}a{\'n}cucki.
\newblock Fastpitch: Parallel text-to-speech with pitch prediction.
\newblock In \emph{ICASSP 2021-2021 IEEE International Conference on Acoustics,
  Speech and Signal Processing (ICASSP)}, pp.\  6588--6592. IEEE, 2021.

\bibitem[Mauch \& Dixon(2014)Mauch and Dixon]{mauch2014pyin}
Matthias Mauch and Simon Dixon.
\newblock pyin: A fundamental frequency estimator using probabilistic threshold
  distributions.
\newblock In \emph{2014 ieee international conference on acoustics, speech and
  signal processing (icassp)}, pp.\  659--663. IEEE, 2014.

\bibitem[McCree \& Barnwell(1995)McCree and Barnwell]{mccree1995mixed}
A.V. McCree and T.P. Barnwell.
\newblock A mixed excitation lpc vocoder model for low bit rate speech coding.
\newblock \emph{IEEE Transactions on Speech and Audio Processing}, 3\penalty0
  (4):\penalty0 242--250, 1995.
\newblock \doi{10.1109/89.397089}.

\bibitem[{Miao} et~al.(2020){Miao}, {Liang}, {Chen}, {Ma}, {Wang}, and
  {Xiao}]{miao20flowtts}
C.~{Miao}, S.~{Liang}, M.~{Chen}, J.~{Ma}, S.~{Wang}, and J.~{Xiao}.
\newblock Flow-tts: A non-autoregressive network for text to speech based on
  flow.
\newblock In \emph{IEEE International Conference on Acoustics, Speech and
  Signal Processing (ICASSP)}, 2020.

\bibitem[Moog(1986)]{moog1986midi}
Robert~A Moog.
\newblock Midi: musical instrument digital interface.
\newblock \emph{Journal of the Audio Engineering Society}, 34\penalty0
  (5):\penalty0 394--404, 1986.

\bibitem[M{\"u}ller et~al.(2019)M{\"u}ller, McWilliams, Rousselle, Gross, and
  Nov{\'a}k]{muller2019neural}
Thomas M{\"u}ller, Brian McWilliams, Fabrice Rousselle, Markus Gross, and Jan
  Nov{\'a}k.
\newblock Neural importance sampling.
\newblock \emph{ACM Transactions on Graphics (TOG)}, 38\penalty0 (5):\penalty0
  1--19, 2019.

\bibitem[Nakatani et~al.(2008)Nakatani, Amano, Irino, Ishizuka, and
  Kondo]{Nakatani2008}
Tomohiro Nakatani, Shigeaki Amano, Toshio Irino, Kentaro Ishizuka, and Tadahisa
  Kondo.
\newblock A method for fundamental frequency estimation and voicing decision:
  Application to infant utterances recorded in real acoustical environments.
\newblock \emph{Speech Communication}, 50\penalty0 (3):\penalty0 203--214,
  March 2008.
\newblock ISSN 0167-6393.
\newblock \doi{10.1016/j.specom.2007.09.003}.
\newblock URL \url{http://dx.doi.org/10.1016/j.specom.2007.09.003}.

\bibitem[Papamakarios et~al.(2017)Papamakarios, Pavlakou, and
  Murray]{papamakariosNeurips17}
George Papamakarios, Theo Pavlakou, and Iain Murray.
\newblock Masked autoregressive flow for density estimation.
\newblock In \emph{Proceedings of the 31st International Conference on Neural
  Information Processing Systems}, NIPS'17, pp.\  2335–2344, Red Hook, NY,
  USA, 2017. Curran Associates Inc.
\newblock ISBN 9781510860964.

\bibitem[Ping et~al.(2020)Ping, Peng, Zhao, and Song]{ping2020waveflow}
Wei Ping, Kainan Peng, Kexin Zhao, and Zhao Song.
\newblock Waveflow: A compact flow-based model for raw audio.
\newblock In \emph{International Conference on Machine Learning}, pp.\
  7706--7716. PMLR, 2020.

\bibitem[Popov et~al.(2021)Popov, Vovk, Gogoryan, Sadekova, and
  Kudinov]{popov2021grad}
Vadim Popov, Ivan Vovk, Vladimir Gogoryan, Tasnima Sadekova, and Mikhail
  Kudinov.
\newblock Grad-tts: A diffusion probabilistic model for text-to-speech.
\newblock \emph{arXiv preprint arXiv:2105.06337}, 2021.

\bibitem[Ren et~al.(2020)Ren, Hu, Tan, Qin, Zhao, Zhao, and
  Liu]{ren2020fastspeech}
Yi~Ren, Chenxu Hu, Xu~Tan, Tao Qin, Sheng Zhao, Zhou Zhao, and Tie-Yan Liu.
\newblock Fastspeech 2: Fast and high-quality end-to-end text to speech.
\newblock \emph{arXiv preprint arXiv:2006.04558}, 2020.

\bibitem[Shih et~al.(2021)Shih, Valle, Badlani, Lancucki, Ping, and
  Catanzaro]{shih2021rad}
Kevin~J Shih, Rafael Valle, Rohan Badlani, Adrian Lancucki, Wei Ping, and Bryan
  Catanzaro.
\newblock Rad-tts: Parallel flow-based tts with robust alignment learning and
  diverse synthesis.
\newblock In \emph{ICML Workshop on Invertible Neural Networks, Normalizing
  Flows, and Explicit Likelihood Models}, 2021.

\bibitem[Suni et~al.(2013)Suni, Aalto, Raitio, Alku, and Vainio]{cwtsuni}
Antti Suni, Daniel Aalto, Tuomo Raitio, Paavo Alku, and Martti Vainio.
\newblock Wavelets for intonation modeling in hmm speech synthesis.
\newblock 01 2013.

\bibitem[Valle et~al.(2020{\natexlab{a}})Valle, Li, Prenger, and
  Catanzaro]{valle2020mellotron}
Rafael Valle, Jason Li, Ryan Prenger, and Bryan Catanzaro.
\newblock Mellotron: Multispeaker expressive voice synthesis by conditioning on
  rhythm, pitch and global style tokens.
\newblock In \emph{ICASSP 2020-2020 IEEE International Conference on Acoustics,
  Speech and Signal Processing (ICASSP)}, pp.\  6189--6193. IEEE,
  2020{\natexlab{a}}.

\bibitem[Valle et~al.(2020{\natexlab{b}})Valle, Shih, Prenger, and
  Catanzaro]{valle2020flowtron}
Rafael Valle, Kevin~J Shih, Ryan Prenger, and Bryan Catanzaro.
\newblock Flowtron: an autoregressive flow-based generative network for
  text-to-speech synthesis.
\newblock In \emph{International Conference on Learning Representations},
  2020{\natexlab{b}}.

\bibitem[Wang et~al.(2017)Wang, Takaki, and Yamagishi]{wangMDN}
Xin Wang, Shinji Takaki, and Junichi Yamagishi.
\newblock {An Autoregressive Recurrent Mixture Density Network for Parametric
  Speech Synthesis}.
\newblock \emph{2017 IEEE International Conference on Acoustics, Speech and
  Signal Processing (ICASSP)}, pp.\  4895--4899, 2017.
\newblock \doi{10.1109/icassp.2017.7953087}.

\bibitem[Wang et~al.(2018{\natexlab{a}})Wang, Takaki, and Yamagishi]{wangDAR}
Xin Wang, Shinji Takaki, and Junichi Yamagishi.
\newblock {Autoregressive Neural F0 Model for Statistical Parametric Speech
  Synthesis}.
\newblock \emph{IEEE/ACM Transactions on Audio, Speech, and Language
  Processing}, 26\penalty0 (8):\penalty0 1406--1419, 2018{\natexlab{a}}.
\newblock ISSN 2329-9290.
\newblock \doi{10.1109/taslp.2018.2828650}.

\bibitem[Wang et~al.(2018{\natexlab{b}})Wang, Takaki, Yamagishi, King, and
  Tokuda]{wangVQDAR}
Xin Wang, Shinji Takaki, Junichi Yamagishi, Simon King, and Keiichi Tokuda.
\newblock {A Vector Quantized Variational Autoencoder (VQ-VAE) Autoregressive
  Neural \$F\_0\$ Model for Statistical Parametric Speech Synthesis}.
\newblock \emph{IEEE/ACM Transactions on Audio, Speech, and Language
  Processing}, 2018{\natexlab{b}}.
\newblock ISSN 2329-9290.
\newblock \doi{10.1109/taslp.2019.2950099}.

\end{thebibliography}
\bibliographystyle{iclr2022}

\newpage
\appendix
\onecolumn

\section{Acoustic Feature Fidelity}\label{appendix:acousticfidelity}
We use acoustic feature fidelity to denote how well a speech synthesis model preserves $F_0$ and $\mathcal{E}$ as specified by the conditioning values. This is important because it describes how speech synthesis models interfere with the conditioning variables and quantifies how much control in fact is available to the end-user.

We evaluate the acoustic fidelity of different models by deconstructing LJSpeech's validation audio files into their extracted phoneme durations, $F_0$, and $\mathcal{E}$, and attempt to reconstruct the original using the RADTTS, FP, and FS2 decoders followed by the Hifi-Gan vocoder. 

We provide quantitative results that compare voiced $F_{0}$ Relative Frame Error (FRE) \cite{kawahara2005nearly}, Voicing Decision Error (VDE) \cite{Nakatani2008}, Energy Error (ENR), and Mel-Spectrogram Error (MER). In addition to results for FP, FS2, the RADTTS decoders, we also measure the irreducible error from the HG vocoder by use of ground-truth mel-spectrograms. This serves as a lower bound to the reconstruction error. All $F_0$ and voicing decisions are extracted using the pYIN algorithm \cite{mauch2014pyin}.


The results in Table~\ref{tab:ffe} show that the RADTTS (RAD) decoder has the lowest reconstruction error in all acoustic features measured. Our results show that RAD performs at least twice as well with respect to $F_0$ fidelity, a significant improvement specially in artistic endeavours where pitch fidelity is of utmost importance.\footnote{In musical terms, 0.05 FRE is equivalent to singing half-a-step out of tune, turning a professional singer into an amateur.}

\begin{figure}[h]
    \begin{subfigure}[tb]{.5\textwidth}
        \caption{Pitch}
        \label{fig:sigma_sweep_pitch}
        \begin{tikzpicture}[scale=1.0]
  \begin{axis}[
  height=4cm,
  width=0.95\linewidth,
  scatter/classes={a={mark=o,draw=brightube}},
       xlabel= {Sigma (std)},
       ylabel= {Error},
      legend cell align={left},
      ylabel near ticks,
      xlabel near ticks,
      xlabel style={font=\small},
      ylabel style={font=\small},
      yticklabel style = {font=\tiny},
      xticklabel style = {font=\tiny},
        y tick label style={/pgf/number format/fixed,/pgf/number format/precision=3},
      xmin=0,
      xmax=1.0,
      ymin=0.018,
      extra y ticks={0.0194}, 
      extra y tick style={
            yticklabel pos=left,
            yticklabels={HifiGan}, 
            ymajorgrids=true  
        }
    ]
    \addplot [scatter, only marks, color=forestgreen] table [x=model, y=f0_rel, col sep=comma] {\sigmasweepdata};
     \addplot[color=red, thick] coordinates {
	(0, 0.0194)
	(1, 0.0194)};
  \end{axis}
\end{tikzpicture}
     \end{subfigure}\hfill
 \begin{subfigure}[tb]{.5\textwidth}
  \caption{Energy}
    \label{fig:sigma_sweep_energy}
    \begin{tikzpicture}[scale=1.0]
  \begin{axis}[
  height=4cm,
  width=0.95\linewidth,
  scatter/classes={a={mark=o,draw=lava}},
       xlabel= {Sigma (std)},
       ylabel= {Error},
      legend cell align={left},
      ylabel near ticks,
      xlabel near ticks,
      xlabel style={font=\small},
      ylabel style={font=\small},
      yticklabel style = {font=\tiny},
      xticklabel style = {font=\tiny},
        y tick label style={/pgf/number format/fixed,/pgf/number format/precision=3},
      xmin=0,
      xmax=1.0,
      ymin=0.040,
      ymax=0.080,
      extra y ticks={0.045}, 
      extra y tick style={
            yticklabel pos=left,
            yticklabels={HifiGan}, 
            ymajorgrids=true  
        }
    ]
    \addplot [scatter, only marks, color=forestgreen] table [x=model, y=energy, col sep=comma] {\sigmasweepdata};
     \addplot[color=red, thick] coordinates {
	(0, 0.045)
	(1, 0.045)};
  \end{axis}
\end{tikzpicture}
 \end{subfigure}
  \begin{subfigure}[tb]{.5\textwidth}
      \caption{Voiced}
    \label{fig:sigma_sweep_voiced}
    \begin{tikzpicture}[scale=1.0]
  \begin{axis}[
  height=4cm,
  width=0.95\linewidth,
       xlabel= {Sigma (std)},
       ylabel= {Error},
      scatter/classes={a={mark=o,draw=dark}},
      legend cell align={left},
      ylabel near ticks,
      xlabel near ticks,
      xlabel style={font=\small},
      ylabel style={font=\small},
      yticklabel style = {font=\tiny},
      xticklabel style = {font=\tiny},
      y tick label style={/pgf/number format/fixed,/pgf/number format/precision=3},
      xmin=0,
      xmax=1.0,
      ymin=0.04,
      extra y ticks={0.044}, 
      extra y tick style={
            yticklabel pos=left,
            yticklabels={HifiGan}, 
            ymajorgrids=true  
        }
    ]
    \addplot [scatter, only marks, color=forestgreen] table [x=model, y=voiced, col sep=comma] {\sigmasweepdata};
     \addplot[color=red, thick] coordinates {
	(0, 0.044)
	(1, 0.044)};
  \end{axis}
\end{tikzpicture}
 \end{subfigure} 
 \begin{subfigure}[tb]{.5\textwidth}
  \caption{MER}
    \label{fig:sigma_sweep_mspc}
    \begin{tikzpicture}[scale=1.0]
  \begin{axis}[
  height=4cm,
  width=0.98\linewidth,
  scatter/classes={a={mark=o,draw=forestgreen}},
       xlabel= {Sigma (std)},
       ylabel= {Error},
      legend cell align={left},
      ylabel near ticks,
      xlabel near ticks,
      xlabel style={font=\small},
      ylabel style={font=\small},
      yticklabel style = {font=\tiny},
      xticklabel style = {font=\tiny},
        y tick label style={/pgf/number format/fixed,/pgf/number format/precision=3},
      xmin=0,
      xmax=1.0,
      ymin=10,
      extra y ticks={12.600}, 
      extra y tick style={
            yticklabel pos=left,
            yticklabels={HifiGan}, 
            ymajorgrids=true  
        }
    ]
    \addplot [scatter, only marks, color=forestgreen] table [x=model, y=mspc, col sep=comma]
    {\sigmasweepdata};
     \addplot[color=red, thick] coordinates {
	(0, 12.6)
	(1, 12.6)};
  \end{axis}
\end{tikzpicture}
 \end{subfigure}
 \caption{Effect of sampling variance on MSE for the RADTTS decoder. The red line in each graph represents
 the irreducible error from the subsequent Hifi-GAN step. While lower sigma values result in lower error, this process also introduces noticeable audio artifacts. As such, we use $\sigma=0.8$ for the purposes of this study.}
 \label{fig:sigma_sweep}
\end{figure}
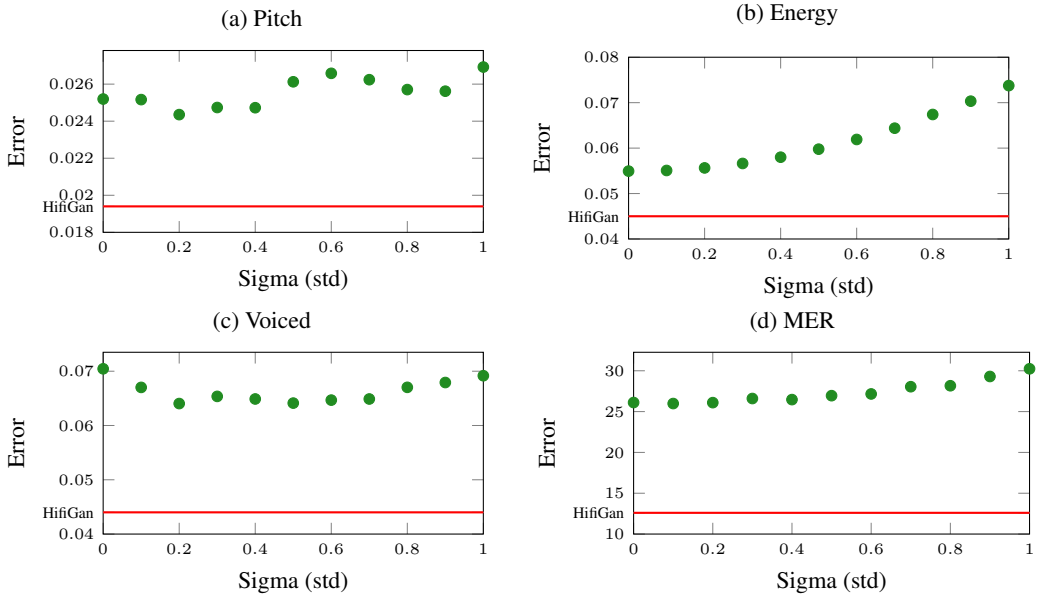

Finally, we include a parameter sweep study of the effect of the sampling noise level on acoustic feature fidelity in Fig.\ref{fig:sigma_sweep}. As observed, several metrics do improve at lower noise levels. However, the lower noise levels also introduce noticeable audio artifacts.

\renewcommand{\arraystretch}{1.1}
\begin{table}[!ht]
\begin{center}
 \caption{The RADTTS decoder has the highest acoustic feature fidelity and lowest mel-spectrogram reconstruction error.}
    \begin{tabular}{ c|c|c|c|c }
        \textbf{Model} &  \textbf{FRE}& \textbf{VDE}& \textbf{ENR}& \textbf{MER}\\
        \hline
        HG  	             & 0.019	& 0.045	& 0.051	& 0.286  \\
        \hline
        \textbf{RADTTS}     	 & \textbf{0.026}	& \textbf{0.067}	& \textbf{0.270}	& \textbf{0.735}  \\        
        FP      	         & 0.055	& 0.128	& 0.355	& 0.833  \\        
        FS2	                 & 0.050	& 0.150	& 0.350	& 0.758 \\

        \hline
    \end{tabular}
    \label{tab:ffe}
\end{center}
\end{table}
\section{Continuous Wavelet Transform implementation details and discussion} \label{sec:cwt}
We follow the exact implementation as used in
FastSpeech2, first filling in the unvoiced areas with linear
interpolation, standardizing the data to be zero mean unit variance,
then applying CWT to give us a 10D representation. Finally, we append
the original mean and variance, replicated along the time axis, to
achieve a 12D representation. This final step is necessary as the mean
and variance are required to undo the standardization step.

There are several drawbacks of this approach. First of all, the CWT is
not a lossless procedure, as the reconstruction has a tendency to
produce a smoother curvature than the original data. Next, while
interpolation is necessary for CWT to be stable, it also means that we
lose track of the location of the unvoiced regions. Furthermore, it is
possible that the interpolated values may introduce noise into the
training data as it falls within the same range as the valid
data. Nevertheless, accounting for these drawbacks, we observe improved training
stability and output quality due to the dimensionality increase as we
will demonstrate later.

\section{Bipartite Model Details}
The bipartite model architecture closely follows that which was used in RADTTS~\cite{shih2021rad}, using steps of flow comprising $1\times1$ invertible convolutions and Coupling transforms. As mentioned in the main text, the proposed architecture replaces 4 out of the 6 steps of flow nearest to the latent space with quadratic splines.

\boldhead{Data Scaling}
Training data should be scaled to be reasonably close in scale to the target standard normal distribution for stability. Centered-difference features are computed on $\log F_0$ and kept at their original scale. $\log F_0$ values are then divided by 6 before being used in the model. $\mathcal{E}$ values are kept at their original values, but the corresponding centered-difference features on $\mathcal{E}$ are multiplied by 10.

\boldhead{Spline Details}
Our work uses quadratic splines in all cases. Previously we have experimented with linear splines, but found quadratic splines to have much better convergence. The splines for the bipartite model operate within the bound $[-3,3]$ with $24$ bins. All values that fall outside of the bounds are passed through with identity, but eventually handled by the $1\times 1$ invertible convolution blocks.

\subsection{Qualitative Analysis}\label{sec:bipartitequal}
\begin{figure}
    \centering
    \includegraphics[width=0.9\linewidth]{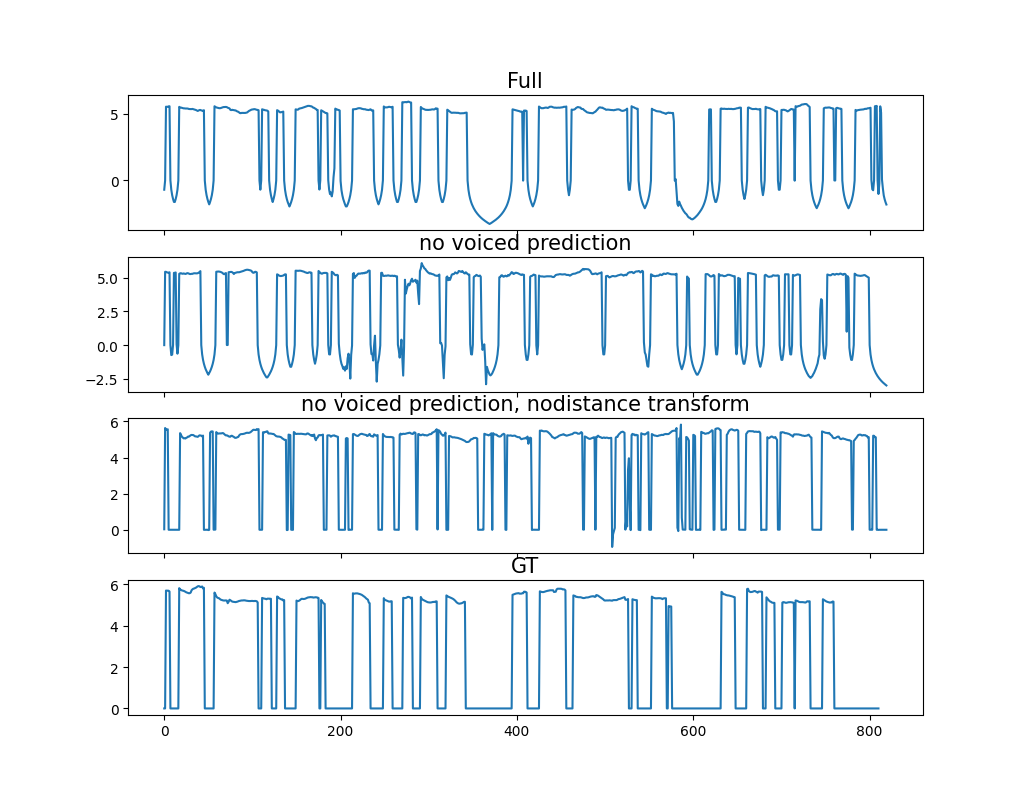}
    \caption{Qualitative ablation of how feature transformations affect $F_0$ prediction quality. The top row is the full model, containing voiced-aware contexts, distance transform filler in the unvoiced regions, and centered-difference features. As voiced prediction and distance transform filler are removed, the model becomes incapable of modeling the unvoiced segments.}
    \label{fig:splinqual}
\end{figure}

Several ablations for the bipartite model were not included in the main studies, as the results were obviously poor. However, we can qualitatively observe the effect of various design choices in Fig.~\ref{fig:splinqual}. We display the direct output of the model in this figure. In other words, if distance transform filler data was used, it should be reproduced by the model during inference, as we see in the top two rows. In practice, these filler values will be easily thresholded away before plugging into the decoder. 

The first thing to observe is the accuracy of how the accuracy of the voiced/unvoiced region segments is affected by the removal of the voiced-aware context. The full model at the top contains voiced-aware contexts, distance transform filler in the unvoiced regions, and centered-difference features. In the second row, we start to see unnatural spikes in the middle of a voiced segment, indicating the model was not certain which mode of behavior to go with at that time step. 

Moving down to the third row, we see how the removal of the distance transform context further degrades the model's ability to model unvoiced regions. Notably, notice that without the distance transform filler data, there are no more long unvoiced segments. We can compare against the ground truth in the fourth row to see that only the full model, and to a lesser extent, the second row with distance transform filler, is able to somewhat accurately model the longer unvoiced segments. However, we note than even in the full model, we can identify false-positive voiced regions at the end of the sample as compared to the ground truth. 

\section{Autoregressive Model Details}\label{sec:appendixagap}
The autoregressive model architecture closely follows that which was used in Flowtron~\cite{valle2020flowtron}. With this setup, it is also possible to reverse the ordering of the attributes in time without loss of generality. We reverse the order of frames on even steps of flow, defining a step of flow as a full pass over the input sequence. This allows the model to learn dependencies both forward and backwards in time while remaining causal and invertible. Our propsed autoregressive flow architecure uses 2 steps of flow, each with 2 LSTMs over the data, the second one also conditioned on the context.

\boldhead{Data Scaling}
Similarly to the Bipartite flow model, training data was scaled to be reasonably close in scale to the target standard normal distribution for stability. 

\boldhead{Spline Details}
Our work uses quadratic splines in all cases. The splines for the autoregressive model operate within the bound $[-6,6]$ with 24 bins. Wider bounds are necessary as we do not incorporate 1x1 invertible convolutions as a catch-all for out-of-bound values.
 
\subsection{Forward and Inverse Procedures}\label{sec:agapfi}
Here, we provide additional background on the forward and inverse procedure for the autoregressive architecture. For simplicity, we will use affine transforms as an example, but the same procedure holds for the neural spline operation. Here, superscripts indicate positional indices, not to be confused with exponents.

Let $X$ be the sequence $x^{0}... x^{T}$. We first augment it with a constant value at the beginning:
\begin{equation}
    X = \mathbf{0}, x^{0}, ..., x^{T}
\end{equation}
Let $NN()$ be the autoregressive transformation parameter predictor that will run through the data from $\mathbf{0}$ to $x^T$. Our first iteration is as follows:
\begin{align}
    s^0, b^0 &= NN(\mathbf{0})\\
    z^0 &= (x^0 - b^0)/s^0
\end{align}
and subsequent steps:
\begin{align}
    s^t, b^t &= NN(\mathbf{0} ... x^{t-1})\\
    z^t &= (x^t - b^t)/s^t
\end{align}

Our end result is a vector $Z = z^{0}...z^T$. To reverse the process, we again start with $\mathbf{0}$ as follows:
\begin{align}
    s^0, b^0 &= NN(\mathbf{0})\\
    x^0 &= s^0z^0 + b^0
\end{align}

and with $x^0$ available, we can then proceed with subsequent steps:
\begin{align}
    s^t, b^t &= NN(\mathbf{0} ... x^{t-1})\\
    x^t &= s^tz^t+b^t
\end{align}

For further information, please see ~\cite{valle2020flowtron} and ~\cite{papamakariosNeurips17}.

\subsection{Qualitative Analysis}
\begin{figure}
    \centering
    \includegraphics[width=\columnwidth]{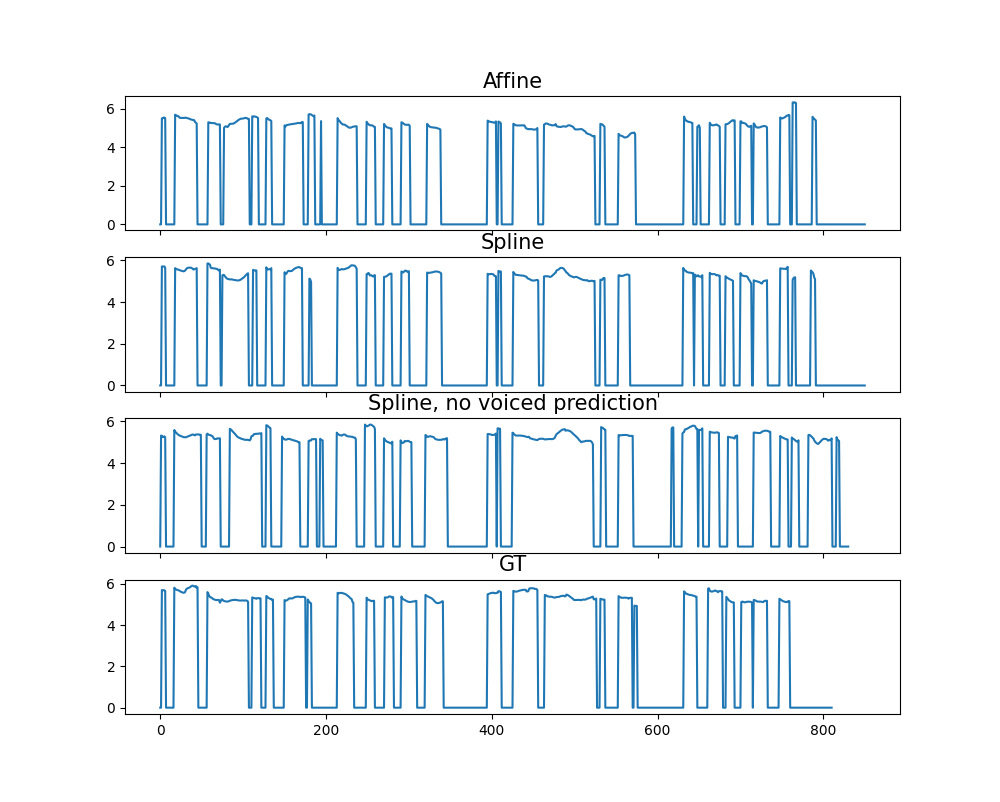}
    \caption{Qualitative analysis of $F_{0}$ synthesis from the autoregressive architecture. While the autoregressive architecture benefits from the use of voiced-aware context, it performs acceptably well compared to a bipartite model without voiced-aware context.}
    \label{fig:agapqual}
\end{figure}

Figure \ref{fig:agapqual} depicts three autoregessive models. The first two use the full set of proposed autoregressive design choices, but one uses affine coupling and the second uses splines.  The qualitative difference between the affine and spline full models is very minimal, supporting the observation that autoregressive are much better suited for the task. In the third row, we have an ablation where voiced-context is removed. We immediately see that there are now more discrepancies between the third row and the ground truth in the fourth. Here, the model must infer voiced-unvoiced state changes based on its internal hidden state, which we suspect is much less reliable than a simple fully-supervised classifier. Nevertheless, we note that the autoregressive result without voiced prediction is still much cleaner than the one from the bipartite model.

\newpage
\section{Pairwise Opinion Scores} 
Subjective scoring is performed by crowd-sourcing pairwise preferences between models, as mean opinion scores are not suited for fine-grained differences. Listeners were pre-screened with a hearing test based on sinusoid counting. Qualified raters were repeatedly given two synthesized utterances of the same text, picked at random from 100 LJ test samples and asked to select samples with best overall quality, defined by accuracy of text, pleasantness, and naturalness. All the model details were hidden from the human raters. Approximately 150 scores per model were collected. Table \ref{tab:pairwise_results} demonstrates the results of the pairwise preference test conducted against specific model combinations. 

First, we observe that the spline flow models are preferred over affine flow based models when sampled with standard deviation $0.5$ as well as $1.0$. The preference is more profound for standard deviation $1.0$. We believe this is a consequence of splines fitting the prior better than affine transforms. 

Next, we compare the models from autoregressive and bipartite flow based generative models against RADDAP, a model with deterministic feature prediction in RADTTS. We observe that in both cases the deterministic feature prediction model is preferred by human raters however there is a confidence interval overlap between the two models. This shows that the generative feature predictors are close in human perception to deterministic feature predictor models. We also observe that the autoregressive spline flow model fares better than the bipartite spline flow model. We also directly compare the autoregressive spline flow model to bipartite spline flow models sampled at standard deviation $0.5$, and observe the same conclusion. The autoregressive feature prediction models outperform parallel feature prediction models and are preferred by human raters.

We compare the speech quality of our generative models using spline and affine transform based flow models against FastPitch and FastSpeech2\footnote{We re-implemented FastSpeech2 to the best of our ability, but due to lack of source code and pretrained models, we had a hard time matching the quality of samples reported by the authors. This comparison was done using our implementation of FastSpeech2.} baselines. We observe that in almost all cases, our deterministic feature prediction model RADDAP as well as generative feature prediction models (spline flow and affine flow models) outperform FastSpeech2 in the perceived speech quality of the samples. In certain cases, the FastPitch model is preferred over our generative feature prediction models but with overlapping confidence intervals. Our deterministic feature prediction model (RADDAP) outperforms both FastPitch and FastSpeech2. Considering all these comparisons, we observe that the RADDAP model is preferred the most among all models considered in this evaluation. However, the spline flow based autoregressive and parallel generative feature prediction models provide a comparable speech synthesis quality with the added benefits of supporting synthesis of diverse samples. 

\begin{table}[!ht]
    \caption{Pairwise preference scores by human raters, shown with $95\%$ confidence intervals of model A (left) vs model B (right). Scores above 0.5 indicate model A was preferred by majority of raters over model B.}
    \vspace*{0.25\baselineskip}
    \centering
    \begin{tabular}{lcc}
        \toprule
        \textbf{Model Pair} & \textbf{Pairwise Preference} & \textbf{Preferred Model} \\
        \midrule
        BGAP Spline ($\sigma=0.5$) vs BGAP Affine ($\sigma=0.5$) &$0.567 \pm 0.0766$ & BGAP Spline\\
        BGAP Spline ($\sigma=1.0$) vs BGAP Affine ($\sigma=1.0$) &$0.647 \pm 0.0774$ & BGAP Spline\\
        \midrule
        RADDAP vs AGAP Spline ($\sigma=0.5$) & $ 0.562 \pm 0.0740 $ & RADDAP (CI Overlap)\\
        RADDAP vs BGAP Spline ($\sigma=0.5$) & $ 0.671 \pm 0.0707 $ & RADDAP\\
        \midrule
        AGAP Spline ($\sigma=0.5$) vs BGAP Spline ($\sigma=0.5$) & $ 0.680 \pm 0.0652 $ & AGAP Spline\\
        \midrule
        BGAP Spline ($\sigma=0.5$) vs FastPitch &$  0.391 \pm 0.0525 $ & FastPitch (CI Overlap)\\
        BGAP Spline ($\sigma=0.5$) vs FastSpeech2 & $0.831 \pm  0.0669$ & BGAP Spline\\
        BGAP Affine ($\sigma=0.5$) vs FastPitch & $0.393 \pm  0.0819 $ & FastPitch\\
        BGAP Affine ($\sigma=0.5$) vs FastSpeech2 & $0.785 \pm  0.0716 $ & BGAP Affine\\
        AGAP Spline ($\sigma=0.5$) vs FastPitch &$  0.537 \pm 0.0688 $ & AGAP Spline (CI Overlap)\\
        AGAP Spline ($\sigma=0.5$) vs FastSpeech2 &$ 0.771 \pm 0.1464 $ & AGAP Spline\\
        RADDAP ($\sigma=0.5$) vs FastPitch &$ 0.586 \pm 0.0692 $ & RADDAP\\
        RADDAP ($\sigma=0.5$) vs FastSpeech2 &$ 0.849 \pm 0.0891 $ & RADDAP\\
        \bottomrule
        \end{tabular}
    \label{tab:pairwise_results}
\end{table}




\end{document}